\documentclass[letterpaper,12pt,hidelinks]{article}
\usepackage{natbib,amsmath,bm,xcolor,amsfonts,graphicx,float,
  subcaption,enumerate,algorithm,algorithmic}
\usepackage[margin=1in]{geometry}

\graphicspath{{figures/}}
\allowdisplaybreaks

\newtheorem{theorem}{Theorem}
\newtheorem{remark}{{\bf Remark}}
\newtheorem{assumption}{\bf Assumption}

\def\be{\begin{eqnarray}}
\def\ee{\end{eqnarray}}
\def\n{{\nonumber}}
\def\wt{\widetilde}

\def\wh{\widehat}
\newcommand{\0}{{0}}
\newcommand{\I}{{I}}
\newcommand{\V}{{V}}
\newcommand{\x}{{x}}
\newcommand{\X}{{X}}

\newcommand{\bbeta}{{\beta}}

\newcommand{\bb}{{\beta}}

\newcommand{\onen}{\frac{1}{n}}
\newcommand{\oneN}{\frac{1}{N}}

\newcommand{\op}{o_P(1)}
\newcommand{\opf}{o_P(1)}
\newcommand{\opfb}{o_P(1)}

\newcommand{\ud}{\mathrm{d}}
\newcommand{\Fn}{\mathcal{F}_N}
\newcommand{\sumn}{\sum_{i=1}^{n}}
\newcommand{\sumN}{\sum_{i=1}^{N}}
\newcommand{\tp}{^{\rm T}}
\newcommand{\tr}{\mathrm{tr}}
\newcommand{\Exp}{\mathbb{E}}
\newcommand{\Nor}{\mathbb{N}}
\newcommand{\Var}{\mathbb{V}}
\newcommand{\mmse}{{\mathrm{Aopt}}}
\newcommand{\mvc}{{\mathrm{Lopt}}}
\newcommand{\opt}{{\mathrm{opt}}}

\begin{document}
\title{Optimal subsampling for quantile regression in big data}
\author{HaiYing Wang\footnote{haiying.wang@uconn.edu}
  \hspace{3mm} and \hspace{3mm}
  Yanyuan Ma\footnote{yzm63@psu.edu}}
\date{\today}
\maketitle
\begin{abstract}
We investigate optimal subsampling for quantile regression. We derive the asymptotic distribution of a general subsampling estimator and then derive two versions of optimal subsampling probabilities. One version minimizes the trace of the asymptotic variance-covariance matrix for a linearly transformed parameter estimator and the other minimizes that of the original parameter estimator. The former does not depend on the densities of the responses given covariates and is easy to implement. Algorithms based on optimal subsampling probabilities are proposed and asymptotic distributions and asymptotic optimality of the resulting estimators are established. Furthermore, we propose an iterative subsampling procedure based on the optimal subsampling probabilities in the linearly transformed parameter estimation which has great scalability to utilize available computational resources. In addition, this procedure yields standard errors for parameter estimators without estimating the densities of the responses given the covariates. We provide numerical examples based on both simulated and real data to illustrate the proposed method.

\end{abstract}

\pagebreak

\section{Introduction}\label{sec:intro}
Quantile regression is an increasingly popular and familiar
tool in statistical analysis. 
Compared with the linear mean regression model, a quantile regression
model has many advantages. For example, it is more robust
so is favored when outliers are present. Quantile regressions at
various quantile levels also provide a more comprehensive picture of
the relation between the response and covariates than the traditional
mean regression, which extracts only the mean relation. In addition,
quantile regression naturally incorporates error heteroscedasticity. 
In big data problems, because data are
often collected from different sources with different times and
locations, the homoscedasticity assumption is often not valid
\citep{fan2014challenges}, which makes quantile regression a natural 
candidate as an analysis tool.

In spite of the aforementioned advantages, it is
computationally difficult to obtain parameter estimates in quantile
regression from
massive data. The simplex algorithm is a popular optimization method for
quantile regression, but it is computationally demanding for large
data sets \citep{chen2005computational}. \cite{portnoy1997gaussian}
introduced the interior point algorithm into quantile regression,
which has been found to be faster than the simplex algorithm when there is
a large 
number of observations. However, the interior point algorithm still
need polynomial time for  optimization; its worst-case time
complexity is $O(N^{5/2}p^3)$,  where $N$ is the sample size and $p$
is the dimension of the regression coefficient \citep[Sec 6.4.4
of][]{koenker2005quantile}. 
Whilst for linear median regression, 
under some conditions, the overall time complexity is
$O(N^{1+a}p^3\log n)$, where $0<a<0.5$ \citep[Theorem 6.3
of][]{koenker2005quantile}. 
In addition, to perform inference through
quantile regression, one often has to rely on the bootstrap method
which further increases the computational burden. This is
because the asymptotic variance-covariance matrix depends on the
densities of the responses given the covariates, which are infeasible to
estimate especially when the dimension of the covariate is high.  

Subsampling has been widely used to reduce
 computational burden when handling massive data.
It performs analysis on a small
subsample drawn from the full data and provides a practical solution to
extracting information from massive data with limited computing
power. This idea has attracted much attention with  extensive
literature such as \cite{
  Drineas:12,  
  dhillon2013new,
  yang2013quantile,
  PingMa2014-JMLR,
  WangZhuMa2018}. Most
existing work takes an algorithmic approach  
and focuses on fast calculation. The first studies to consider
statistical properties include \cite{PingMa2014-JMLR,
  raskutti2014statistical}  and \cite{WangZhuMa2018}. Specifically, 
\cite{PingMa2014-JMLR}
assessed biases and variances for subsampling estimators based on
statistical leverage scores in linear regression;
\cite{raskutti2014statistical} investigated ordinary least-squares
estimators based on randomized sketching; and \cite{WangZhuMa2018}
proposed an optimal subsampling method under the A-optimality
criterion for logistic regression. \cite{wang2018more}
  proposed a more efficient estimator based on the optimal subsample,
  and \cite{ai2018optimal} extended the optimal subsampling technique
  to generalized linear models. \cite{wang2019information} proposed a
  method called information-based optimal subdata selection for linear
  mean regression, which selects subsamples deterministically without
  involving random sampling.

In this paper, we use the idea of optimal subsampling to meet the
challenges in computation and inference for quantile regression.  
We derive the asymptotic distribution of a general subsampling based
estimator and find the optimal subsampling probabilities that minimize a
weighted version of the asymptotic mean squared errors (MSE). In
addition to the computational advantage, the subsampling technique
also provides a scalable approach to perform statistical inference. 
The theory of optimal subsampling cannot be easily extended to
  quantile regression, because it only applies
when the target function is smooth and at least twice
  differentiable, which is not satisfied in the quantile regression context.
Compared with standard practices for quantile regression, the asymptotic results are
  significantly more challenging to obtain in the context of
  subsampling. There are two layers of randomness for a subsample, one
  is from the randomness of the data and the other is due to
  subsampling. Both sources of the randomness need to be taken into
  account in the proof. In addition, although the subsample
  observations are independent conditional on the full data, they are
  correlated unconditionally, which further complicates the
  analysis. In this paper, we do not consider the 
  deterministic selection method in \cite{wang2019information},
  because this method requires to characterize the exact
  variance-covariance matrix of the subsample estimator which is not
  feasible for quantile regression.  

An alternative popular approach to dealing with massive data is the
  divide and conquer method that first divides the full data into
  small pieces to analyze, and then combines the analysis results from
  all pieces to obtain an aggregated estimator. More details about
  this approach can be found in
  \cite{LinXie2011,schifano2016online,cheng2015computational,volgushev2017distributed}  
  and the references therein. This approach mainly aims at analyzing
  the full data with parallel or distributed computing platform, while
  the subsampling method aims at fast calculation with limited
  computing resources. 


\section{Problem Statement}
\label{sec:problem-statement}

\subsection{Model}\label{sec:model}
Consider a linear quantile regression model 
\begin{equation}\label{eq:22}
q_\tau(Y_i\mid\x_i)=\bbeta\tp\x_i,
\end{equation}
where $q_\tau(Y_i\mid\x_i)$ is the $\tau$-th 
quantile of the univariate response $Y_i$ at a given value of
  the $p$-dimensional covariate 
vector $x_i$. In this paper, we assume that $x_i$'s are nonrandom, and
we want to estimate the unknown $\bbeta$ from observed 
data of size $N$, $(\x_i,y_i), i=1, \dots, N$,
where the
true $\bbeta$ value is assumed to be in the interior of a
compact set. 

\subsection{Full data estimation of $\bbeta$}
Let $\varepsilon_i=y_i-\bbeta\tp\x_i$, and let
$f_{\varepsilon\mid\X}(\varepsilon_i,\x_i)$ be the probability density
function of $\varepsilon_i$ evaluated at 
$\varepsilon_i$ with covariate $\x_i$. The most frequently seen method
of estimating of $\bbeta$ is through minimizing 
\begin{align}\label{eq:24}
    Q_N(\bbeta)=\oneN\sumN\rho_\tau(y_i-\bbeta\tp\x_i)
     =\oneN\sumN(y_i-\bbeta\tp\x_i)\{\tau-I(y_i<\bbeta\tp\x_i)\},
\end{align}
where $\rho_\tau(\cdot)$ is the check function defined as
$\rho_\tau(\varepsilon)=\varepsilon\{I(\varepsilon\ge0)-(1-\tau)\}
=\varepsilon\{\tau-I(\varepsilon<0)\}$.   

Denote the minimizer of \eqref{eq:24} as $\wh\bbeta$. Under some
regularity conditions, the full data estimator $\wh\bbeta$ has some
desirable asymptotic properties. Here we adopt the set of regularity
conditions used in \cite{koenker2005quantile} and list them below as
Assumption~\ref{asp1} for completeness.  

\begin{assumption}\label{asp1}\-\\[-7mm]
  \begin{enumerate}[(a)]
  \item Assume that $f_{\varepsilon\mid\X}(t,\x)$ is continuous with
    respect to $t$ and is uniformly bounded away from 0 and $\infty$
    at $t=0$.
  \item Assume that there exist positive definite matrices $D_0$ and $D$ such that
    \begin{align}
      D_{N0}\equiv&\oneN\sumN\x_i\x_i\tp\rightarrow D_0,\label{eq:41}\\
      D_N\equiv&\oneN\sumN f_{\varepsilon\mid\X}(0,\x_i)\x_i\x_i\tp
            \rightarrow D,\label{eq:42}\\
      &\frac{\max_{1\le i\le N}\|\x_i\|}{\sqrt{N}}=o(1).\label{eq:43}
    \end{align}
  \end{enumerate}
\end{assumption}

As shown in Theorem 4.1 of \cite{koenker2005quantile}, under
Assumption~\ref{asp1}, the full data estimator $\wh\bbeta$ satisfies
that
\begin{align}\label{eq:21}
  \{\tau(1-\tau)D_N^{-1}D_{N0}D_N^{-1}\}^{-1/2}\sqrt{N}
  (\wh\bbeta-\bbeta_t)\longrightarrow\Nor(\0, \I),
\end{align}
in distribution, 
where $\Nor(\0, \I)$ represents a multivariate standard normal
distribution, and $\bbeta_t$ stands for the true value of $\bbeta$. 
This  result indicates that the distribution of
$\wh\bbeta$ can be approximated by a normal
distribution for large $N$, and this forms the basis for statistical
inference on $\bbeta$ or on the quantile of the
response given the covariates. However, for massive data with very large $N$, it is
computationally difficult to obtain $\wh\bbeta$ numerically. In
addition, \eqref{eq:21} is often not usable for statistical inference
because it is hard to obtain estimates of $f_{\varepsilon\mid\X}(0,\x_i)$ in
the expression of $D_N$. To solve these issues and to apply quantile
regression for massive data, we develop a subsampling based approach in
the following sections.

\section{Subsampling based estimation}
\label{sec:subs-based-estim}
\subsection{Subsampling based estimator and its asymptotic
  distribution}\label{sec:31}

Take a random subsample using sampling with replacement from the full
data according to the probabilities $\pi_i$, $i=1,...N$, such that
$\sumN\pi_i=1$. Here $\pi_i$ may depend on the full data
$\Fn=\{(\x_i,y_i), i=1, \dots, N\}$. In this paper, we use
  sampling with replacement because nonuniform sampling without
  replacement 
  requires to update the sampling distribution sequentially based on
  selected observations (e.g., in the \texttt{sample} function of R),
  which is computationally slow. In addition, when the sampling ratio
  is very small, sampling with and without replacement have very
  similar performance. Denote the subsample as 
$(\x^{*}_i, y^{*}_i)$, with associated subsampling probabilities
$\pi_i^*$, $i=1,...,n$. The subsample estimator, denoted as
$\wt\bbeta$, is the minimizer of  
\begin{equation}\label{eq:12}
  Q_n^*(\bbeta)=
  \onen\sumn\frac{\rho_\tau(y_i^*-\bbeta\tp\x_i^*)}{N\pi_i^*},
\end{equation}
which can be equivalently written as
\begin{eqnarray*}
Q_n^*(\bbeta)=\frac{1}{nN}\sumN\frac{R_i\rho_\tau(y_i-\bbeta\tp\x_i)}{\pi_i},
\end{eqnarray*}
where $R_i$ is the total number of times that the $i$th observation is
selected into the sample out of the $n$ sampling steps. 
Here, we need to weight the target function based on
  the subsampling probabilities $\pi_i^*$'s, because we allow
  $\pi_i$'s to depend on the responses $y_i$'s and an un-weighted
  target function would result in a biased estimator.

We now show the asymptotic normality of $\wt\bb$, and then identify the
 $\pi\equiv\{\pi_1, ..., \pi_N\}$ that minimizes the asymptotic variance. To establish the
asymptotic normality, we assume some conditions on the subsampling
probabilities  in Assumption~\ref{asp2}. Note that we allow $\pi_i$'s
to be dependent on the responses $y_i$'s, so they may be random.

\begin{assumption}\label{asp2}\-\\[-7mm]
  \begin{enumerate}[(a)]
  \item Assume that
    \begin{equation}
      \max_{1\le i\le N}\frac{\|\x_i\|}{\pi_i}=o_P(\sqrt{n}N).\label{eq:17}
    \end{equation}
\item Assume that 
  \begin{equation}
    \V_\pi=\sumN\frac{\{\tau-I(\varepsilon_i<0)\}^2\x_i\x_i\tp}{N^2\pi_i}\label{eq:vc}
  \end{equation}
  converges to a positive definite matrix in probability.
\end{enumerate}
\end{assumption}

\begin{remark}
 Assumption 2 contains two requirements on the
 sampling probabilities $\pi_i$'s. 
These are not very restrictive conditions as one can see by inserting
equal probabilities $\pi_i=1/N$. They mainly require that the 
maximum 
covariates 
weighted by the inverse selecting probabilities do not diverge or diverge
too fast.
\end{remark}

The following theorem describes the asymptotic normality of
$\wt\bbeta$.

\begin{theorem}\label{thm:1}
  Under Assumptions 1 and 2, as $n\rightarrow\infty$ 
and
  $N\rightarrow\infty$, if $n=o(N)$, then
  $\sqrt{n}(\wt\bbeta-\bbeta_t)$ asymptotically follows a normal
  distribution with mean $\0$ and variance-covariance matrix 
  approximated by $D_N^{-1}\V_\pi D_N^{-1}$, i.e. 
  \begin{equation*}
    (D_N^{-1}\V_\pi D_N^{-1})^{-1/2}
    \sqrt{n}(\wt\bbeta-\bbeta_t)
    \longrightarrow\Nor(\0, \I)
  \end{equation*}
  in distribution, 
  where $D_N$ is defined in (\ref{eq:42}) and $\V_\pi $ is defined in (\ref{eq:vc}). 
\end{theorem}

\subsection{Optimal subsampling probabilities}

The asymptotic distribution of $\wt\bbeta$ depends on the subsampling
probabilities $\pi_i$'s and the key to the success of a subsampling
based estimator is to find the $\pi_i$'s to optimize some criterion
of the asymptotic distribution. Since $\wt\bbeta$ is asymptotically
unbiased, we focus on minimizing the asymptotic variance-covariance matrix. 

In the asymptotic variance-covariance matrix $n^{-1}D_N^{-1}\V_\pi
D_N^{-1}$, only $\V_\pi$ depends on $\pi_i$'s while $D_N$ does not
involve $\pi_i$'s, and $D_N^{-1}\V_{\pi} D_N^{-1}\le D_N^{-1}\V_{\pi'}
D_N^{-1}$ if and only if $\V_{\pi}\le \V_{\pi'}$ in the Loewner
ordering \citep{Yang2010}. 
In addition, $D_N$ depends on the 
density functions of
$\varepsilon_i$'s at zero given the respective $\x_i$'s, which are often infeasible to estimate in
practice. Thus, we 
propose to focus on minimizing $\V_{\pi}$. 
As there is no complete ordering for matrices, a natural choice is to
minimize the trace. Therefore, we propose to find optimal subsampling
probabilities to minimize $\tr(\V_\pi)$. 
Note that $n^{-1}\V_\pi$ can be viewed as the asymptotic
variance-covariance matrix of $D_N\wt\bbeta$ in estimating $D\bbeta$,
a linearly transformed parameter. Thus, minimizing $\tr(\V_\pi)$ can
be interpreted as minimizing the asymptotic MSE of $D_N\wt\bbeta$ due
to its asymptotic unbiasedness. This choice also has an optimality
interpretation in terms of optimal experimental design; it is
termed the L-optimality criterion, where ``L'' stands for ``linear
transformation'' of the estimator 
\citep[see][]{Atkinson2007Optimum}. Using this criterion we are able to
obtain the explicit expression of optimal subsampling probabilities in
the following theorem. 

\begin{theorem}[L-optimality]\label{thm:3}
  If the sampling probabilities $\pi_i$, $i=1,...N$, are chosen as
  \be
    \pi_i^{\mvc}=
    \frac{|\tau-I(\varepsilon_i<0)|\|\x_i\|}
    {\sum_{j=1}^N|\tau-I(\varepsilon_j<0)|\|\x_j\|},\;
    i=1,2,...,N,\label{eq:8}
  \ee
  then the total asymptotic MSE of $D_N\wt\bbeta$, $\tr(\V_\pi )/n$, attains its minimum.
\end{theorem}

For completeness, we also derive the optimal subsampling
probabilities that minimize the asymptotic MSE of $\wt\bbeta$, that
is, the $\pi_i$'s that minimize the trace of $n^{-1}D_N^{-1}\V_\pi D_N^{-1}$. This is
called the A-optimality criterion in optimal experimental design
\citep[see][]{Atkinson2007Optimum}.  

\begin{theorem}[A-optimality]\label{thm:2}
  If the sampling probabilities $\pi_i$, $i=1,\dots, N$ are chosen as
  \begin{eqnarray*}
    \pi_i^{\mmse}=
    \frac{|\tau-I(\varepsilon_i<0)|\|D_N^{-1}\x_i\|}
    {\sum_{j=1}^N|\tau-I(\varepsilon_j<0)|\|D_N^{-1}\x_j\|},\;
    i=1,2,...,N,\label{eq:7}
  \end{eqnarray*}
  then the total asymptotic MSE of $\wt\bbeta$, $\tr(D_N^{-1}\V_\pi D_N^{-1})/n$, attains its
  minimum.
\end{theorem}

\begin{remark}
  The L-optimal subsampling probabilities $\pi_i^{\mvc}$'s do not depend on the 
 densities of $\varepsilon_i$'s given the associated $\x_i$'s and thus are much easier to
implement compared with the A-optimal subsampling probabilities
$\pi_i^{\mmse}$'s, which depend on the conditional density through
$D_N$. In addition, $\pi_i^{\mvc}$'s require $O(Np)$  
time to compute, while $\pi_i^{\mmse}$ require $O(Np^2)$ time to
compute even if $D_N$ is available.
\end{remark} 

In \eqref{eq:8},  $\varepsilon_i=y_i-\bbeta\tp\x_i$, and it
depends on the unknown $\bbeta$, so the L-optimal weight result is not
directly implementable.  We propose the following two-step
algorithm to address this issue.  

\begin{algorithm}[H]
  \caption{Two-step Algorithm in implementing $\pi_i^{\mvc}$}
  \label{alg:2}
  \begin{itemize}
  \item \textbf{Step 1:} Using the uniform sampling probability
    $\pi_i^{0}=1/N$, draw a random subsample
    of size $n_0$ to obtain a preliminary estimate of $\bbeta$,
    $\wt\bbeta_0$. Replace $\bbeta$ with $\wt\bbeta_0$ in \eqref{eq:8}
    to obtain the approximate optimal subsampling probabilities
    $\pi_i^{\mvc,\wt\bb_0}$.

  \item \textbf{Step 2:} Subsample with replacement to obtain a
    subsample of size $n$ using $\pi_i^{\mvc,\wt\bb_0}$, and use it to 
    obtain the estimate
    $\breve{\bbeta}_{\mvc}$ 
    through minimizing
    \begin{equation}\label{eq:9}
      Q_n^{*(2)}= \frac{1}{n}\sumn
      \frac{\rho_\tau(y_i^*-\bbeta\tp\x_i^*)}{N\pi_i^{*\mvc,\wt\bbeta_0}}.
    \end{equation}
  \end{itemize}
\end{algorithm}

If the density $f_{\varepsilon\mid\X}(0,\x)$ is
obtainable, then $\pi_i^{\mmse}$ can be implemented similarly as in 
Algorithm~\ref{alg:2} to obtain $\breve\bbeta_\mmse$. 
In this case, we can further combine the pilot estimator and the
second step estimator. To be specific, let $\tilde
f_{\varepsilon\mid\X}(0,\x)$ be the estimate of 
$f_{\varepsilon\mid\X}(0,\x)$ based on the first step sample, and let
\begin{eqnarray*}
  \wt{D}_{n_0}=\frac{1}{n_0}\sum_{i=1}^{n_0}\frac
  {\tilde f_{\varepsilon\mid\X}(0,\x_i^{*0})\x_i^{*0}{\x_i^{*0}}\tp}
  {N\pi_i^{*0}}
  \quad\text{ and }\quad
  \wt{D}_{n}=\frac{1}{n}\sumn\frac
  {\tilde f_{\varepsilon\mid\X}(0,\x_i^{*})\x_i^{*}{\x_i^{*}}\tp}
  {N\pi_i^{*\mmse,\wt\bb_0}},
  \end{eqnarray*}
  where $\pi_i^{*0}=1/N$, and $(\x_i^{*0})_{i=1}^{n_0}$ and
$(\x_i^{*})_{i=1}^{n}$ are respectively the first and second step
subsample covariates. 
After obtaining the second step estimator $\breve\bbeta_\mmse$, we can
aggregate it with the pilot estimator 
$\wt\bbeta_0$ 
using
\begin{equation}\label{eq:11}
  (n_0\wt{D}_{n_0}+n\wt{D}_{n})^{-1}n_0\wt{D}_{n_0}\wt\bbeta_0
  +(n_0\wt{D}_{n_0}+n\wt{D}_{n})^{-1}n\wt{D}_{n}\breve\bbeta_\mmse.
\end{equation}
The linear combination in \eqref{eq:11} is similar to the
aggregation step in the divide and conquer method
\citep{LinXie2011,schifano2016online}, and is used to further improve the
estimation variability from $\breve\bbeta_\mmse$.

In practice, with limited computing resources, one often takes a pilot
subsample with size $n_0$ to explore the data, and then select a
second subsample 
with size $n$ according to the computational capacity available. 
It is not recommended to combine the two step subsamples to perform
estimation. This is because if we are willing to handle estimation
under size $n_0+n$, then we could have chosen a better sample by
setting the second step sample size to $n_0+n$ directly. Thus, unless 
$f_{\varepsilon\mid\X}(0,\x)$ is available, in which case we can
further improve our estimation via (\ref{eq:11}), the first step
subsample should only be used to help estimate the second step sampling
weights. It should not participate in the second step estimation directly.

In Algorithm \ref{alg:2}, the
pilot estimate is used to calculate the
approximate optimal subsampling probabilities. We have the
following theorem to describe the asymptotic properties of the resultant 
estimators $\breve{\bbeta}_\mvc$ and $\breve{\bbeta}_\mmse$.

\begin{theorem}\label{thm:4}
  Assume that 
  $N^{-1}\sumN\|\x_i\|^{-1}{\x_i\x_i\tp}$
  converges to a positive definite matrix. 
  Under Assumption 1, as $n_0\rightarrow\infty$, $n\rightarrow\infty$ and $N\rightarrow\infty$, if 
  $n=o(N)$, then 
  the 
  distribution of
  $\sqrt{n}(\breve{\bbeta}_\mvc-\bbeta_t)$ is asymptotically normal, i.e., 
  \begin{equation}\label{eq:15}
    (D_N^{-1}\V_{\mvc}D_N^{-1})^{-1/2}
    \sqrt{n}(\breve\bbeta_{\mvc}-\bbeta_t)\longrightarrow
    \Nor(\0, \I)
  \end{equation}
  in distribution, 
where $\V_{\mvc}$ has the minimum trace, and it has the explicit expression
\begin{equation}\label{eq:46}
  \V_{\mvc}=\oneN\sumN\frac{|\tau-I(\varepsilon_i<0)|\x_i\x_i\tp}{\|\x_i\|}
  \times \oneN\sumN|\tau-I(\varepsilon_i<0)|\|\x_i\|. 
\end{equation}
Furthermore, 
if
$\sup_{\x}|\wt{f}_{\varepsilon\mid\X}(0,\x)-f_{\varepsilon\mid\X}(0,\x)|=\op$,
then  $\sqrt{n}(\breve{\bbeta}_\mmse-\bbeta_t)$ is asymptotically normal, i.e.
\begin{eqnarray*}
(D_N^{-1}\V_{\mmse}D_N^{-1})^{-1/2}
\sqrt{n}(\breve\bbeta_{\mmse}-\bbeta_t)
\longrightarrow
\Nor(\0, \I)
\end{eqnarray*}
in distribution. 
In this case,  
$D_N^{-1}\V_{\mmse}D_N^{-1}$ has the minimum trace,  and $\V_{\mmse}$ 
has the explicit expression
\begin{equation}\label{eq:47}
  \V_{\mmse}=
  \oneN\sumN\frac{|\tau-I(\varepsilon_i<0)|\x_i\x_i\tp}{\|D_N^{-1}\x_i\|}
  \times \oneN\sumN|\tau-I(\varepsilon_i<0)|\|D_N^{-1}\x_i\|.
\end{equation}
\end{theorem}

\section{Iterative subsampling based on $\pi_i^{\mvc}$}
\label{sec:iterative-sampling}

For statistical inference, to avoid estimating
$f_{\varepsilon\mid\X}(0,\x)$, which appears in the asymptotic
variance-covariance matrix expression of $\breve\bbeta_{\mvc}$, we
propose the
following iterative sampling procedure based on $\pi_i^{\mvc}$ that
will produce both the 
point estimator and the standard deviation. Moreover, the convergence
rate of the point estimator is proportional to the square root of the
number of iterations. This provides great scalability for the algorithm
to extract information from big data according to the 
available computing resources.  

\begin{algorithm}[H]
\caption{Two-step iterative sampling algorithm with $\pi_i^{\mvc}$}
  \label{alg:3}
  \begin{itemize}
  \item \textbf{Step 1:} Using the uniform sampling probability
    $\pi_i^{0}=1/N$, draw a random subsample
    of size $n_0$ to obtain a preliminary estimate of $\bbeta$,
    $\wt\bbeta_0$. Replace $\bbeta$ with $\wt\bbeta_0$ in \eqref{eq:8}
    to obtain the approximate optimal subsampling probabilities
    $\pi_i^{\wt\bb_0}$.

  \item \textbf{Step 2:} For $b=1,\dots,B$, subsample with replacement
    to obtain subsamples of size $n$ using $\pi_i^{\wt\bb_0}$, obtain
    $\breve{\bbeta}_{\mvc,b}$ through minimizing
    \begin{eqnarray*}
      Q_n^{*(2)}= \frac{1}{n}\sumn
      \frac{\rho_\tau(y_i^*-\bbeta\tp\x_i^*)}{N\pi_i^{*\wt\bbeta_0}},
    \end{eqnarray*}
    and calculate\\[-6mm]
    \begin{equation}\label{eq:25}
      \breve{\bbeta}_I = \frac{1}{B}\sum_{b=1}^B\breve{\bbeta}_{\mvc,b}
    \end{equation}
    and its variance-covariance estimator
    \begin{equation}\label{eq:23}
      \wh\Var(\breve{\bbeta}_I)=\frac{1}{r_{ef}B(B-1)}
      \sum_{b=1}^B(\breve{\bbeta}_{\mvc,b}-\breve{\bbeta}_I)^{\otimes2},
    \end{equation}
    where\\[-8mm]
    \begin{equation}\label{eq:ref}
      r_{ef}=1-\frac{nB-1}{2}\sumN (\pi_i^{*\wt\bbeta_0})^2.
    \end{equation}
  \end{itemize}
\end{algorithm}

\begin{remark}
  The term $r_{ef}$ is a correction term for effective subsample
  size. Since the subsampling is with replacement, the number of
  unique observations in a subsample may be smaller than $n$. Although
  the probability for this scenario to occur converges to zero if
  $n/N\rightarrow0$, using $r_{ef}$ helps to improve the finite sample
  performance of the variance-covariance estimator. The correction
  term $r_{ef}$ is derived as the following. Given the full data and
  subsampling probabilities, for each observation, the probability that
  it is included in a subsample is 
  \begin{eqnarray*}
    1-(1-\pi_i^{*\wt\bbeta_0})^{nB}
    &\approx&1-\Big\{1-nB\pi_i^{*\wt\bbeta_0}
      +\frac{nB(nB-1)}{2}(\pi_i^{*\wt\bbeta_0})^2\Big\}\\
    &=&nB\pi_i^{*\wt\bbeta_0}
      -\frac{nB(nB-1)}{2}(\pi_i^{*\wt\bbeta_0})^2.
  \end{eqnarray*}
  Thus, the expected effective total subsample size is approximated by 
  \begin{eqnarray*}
    n_{ef}=\sumN \Big\{nB\pi_i^{*\wt\bbeta_0}
      -\frac{nB(nB-1)}{2}(\pi_i^{*\wt\bbeta_0})^2\Big\}
      =nB-\frac{nB(nB-1)}{2}\sumN(\pi_i^{*\wt\bbeta_0})^2. 
  \end{eqnarray*}
  This gives the effective subsample size ratio 
  $r_{ef}=n_{ef}/(nB)$ as given in (\ref{eq:ref}).
\end{remark}
  From Theorem~\ref{thm:4}, for any fixed $B$,  
the conditional distribution of
$\sqrt{nB}(\breve\bbeta_I-\bbeta_t)$ satisfies 
\begin{equation}\label{eq:14}
  (D_N^{-1}\V_{\mvc}D_N^{-1})^{-1/2}\sqrt{nB}(\breve\bbeta_I-\bbeta_t)
  \longrightarrow\Nor(\0, \I).
\end{equation}
To ensure that the bias is ignorable compared to the variance, 
 the result in \eqref{eq:14} requires a fixed $B$ while requiring
$n\to\infty$. 
This indicates that in practice, we should choose $n$
to be as large as it is feasible while select a relatively small $B$. 
An overly large $B$ value can risk 
leading to incorrect inference results. Similar performance is
also observed in the divide and conquer procedures \citep{schifano2016online, cheng2015computational, battey2018distributed,volgushev2017distributed}.
In practice, we find that
often $B$ as small as 10 is already sufficient while it is preferably
$\le n/10$. Please refer to
Section \ref{sec:nB} in the supplement for numerical examples.

Our analysis and optimal sampling probabilities are
  tailored to the specific quantile level $\tau$. In the situaiton
  when we need to consider several, say $M$, quantiles simultaneously, we can
  either perform the analysis for each quantile, or opt for a
  sub-optimal universal approach which is computational simpler.
To this end, note that at a fixed $\tau_m$, 
our $L$ optimal subsampling probabilities minimize 
$ \sumN{\{\tau_m-I(\varepsilon_i<0)\}^2
      \x_i\tp\x_i}/{(N^2\pi_i)},
 $
which is upper bounded by
 $N^{-2}\max\{\tau_m^2,(1-\tau_m)^2\}
    \sumN\x_i\tp\x_i\pi_i^{-1}$.
Hence we can minimize
$\sumN\x_i\tp\x_i\pi_i^{-1}$ to obtain the sub-optimal universal sampling probabilities
    $\pi_i^{U}= {\|\x_i\|}(\sum_{j=1}^N\|\x_j\|)^{-1}$, for
    $i=1,\dots, N$. We conducted additional numerical experiments to
    evaluate the performance of these universal probabilities in
    Section \ref{sec:manytaus} in the supplement, and the efficiency loss does not seem
    to be severe.

\section{Numerical experiments}
\label{sec:numer-exper}

\subsection{Simulation}

We first conduct a simulation study. Full data of size $N=10^6$ are
generated from model~\eqref{eq:22} 
with the true value of $\bbeta$, $\bbeta_t$, being a $7\times1$ vector 
of ones. We consider the following 3 different distributions to generate the covariate $\X$:
\begin{enumerate}[{ }]
\item 1) 
  Multivariate normal distribution 
  $N({0},{\Sigma})$, where
  ${\Sigma}_{ij}=0.5^{|i-j|}$; 
\item 2) Multivariate $t$ distribution with degrees of freedom 3,
  $t_3({0},{\Sigma})$; and
\item 3) Multivariate $t$ distribution
  with degrees of freedom 2, $t_2({0},{\Sigma})$.
\end{enumerate}

  We consider two values of $\tau$: 0.5 and 0.75. For the
  distributions of the response $Y$ given $\X$, we consider three
  different cases:
  \begin{enumerate}[{ }]
  \item    1) the standard normal distribution times
  $(1/7)\sum_{j=1}^7|X_j|$;
\item  2) exponential distribution with rate
  parameter 1 times $(1/7)\sum_{j=1}^7|X_j|$; and
\item 3) $t_1$
  distribution times $(1/7)\sum_{j=1}^7|X_j|$.
\end{enumerate}

  We take $n_0=1000$ and $n=1000$, and calculate MSEs of
  $\breve{\bbeta}_I$ based on $S=1000$ repetitions of the simulation
  using 
$\textrm{MSE}=S^{-1}\sum_{s=1}^S\|\breve{\bbeta}_I^{(s)}-\bbeta_0\|^2$,
where $\breve{\bbeta}_I^{(s)}$ is the estimate from the $s$th
repetition of the simulation.

\begin{figure}[htp]
  \centering
\includegraphics[width=0.2\textwidth,page=1]{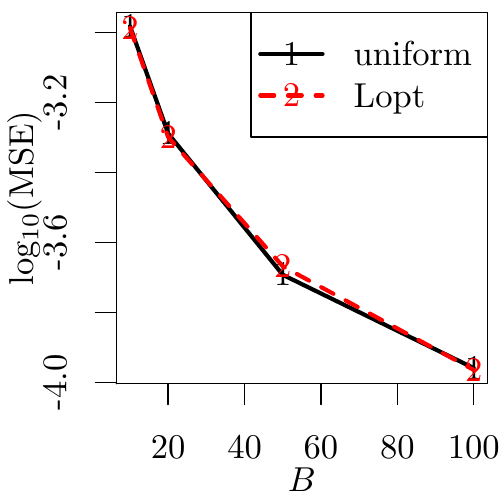}
\includegraphics[width=0.2\textwidth,page=2]{Bootmse.pdf}
\includegraphics[width=0.2\textwidth,page=3]{Bootmse.pdf}
\includegraphics[width=0.2\textwidth,page=4]{Bootmse.pdf}
\includegraphics[width=0.2\textwidth,page=5]{Bootmse.pdf}
\includegraphics[width=0.2\textwidth,page=6]{Bootmse.pdf}
\includegraphics[width=0.2\textwidth,page=7]{Bootmse.pdf}
\includegraphics[width=0.2\textwidth,page=8]{Bootmse.pdf}
\includegraphics[width=0.2\textwidth,page=9]{Bootmse.pdf}
\includegraphics[width=0.2\textwidth,page=10]{Bootmse.pdf}
\includegraphics[width=0.2\textwidth,page=11]{Bootmse.pdf}
\includegraphics[width=0.2\textwidth,page=12]{Bootmse.pdf}
\includegraphics[width=0.2\textwidth,page=13]{Bootmse.pdf}
\includegraphics[width=0.2\textwidth,page=14]{Bootmse.pdf}
\includegraphics[width=0.2\textwidth,page=15]{Bootmse.pdf}
\includegraphics[width=0.2\textwidth,page=16]{Bootmse.pdf}
\includegraphics[width=0.2\textwidth,page=17]{Bootmse.pdf}
\includegraphics[width=0.2\textwidth,page=18]{Bootmse.pdf}
\includegraphics[width=0.2\textwidth,page=19]{Bootmse.pdf}
\includegraphics[width=0.2\textwidth,page=20]{Bootmse.pdf}
\includegraphics[width=0.2\textwidth,page=21]{Bootmse.pdf}
\includegraphics[width=0.2\textwidth,page=22]{Bootmse.pdf}
\includegraphics[width=0.2\textwidth,page=23]{Bootmse.pdf}
\includegraphics[width=0.2\textwidth,page=24]{Bootmse.pdf}
\caption{log$_{10}$(MSE) against number of repeat subsampling $B$. The
  three columns 1-3 correspond to the three distributions of $\X$ (normal, $t_3$, $t_2$), respectively. Rows 1-3 are for $\tau=0.5$ and rows 4-6 are
  for $\tau=0.75$. Rows 1 and 4, 2 and 5, and 3 and 6 are for cases
  when $Y$ follows normal, exponential, and $t_1$ distributions,
  respectively.} 
  \label{fig:01}
\end{figure}

Figure~\ref{fig:01} presents MSEs for different scenarios using
$\pi_i^{\mvc}$. For better
presentation, we show MSEs on the
$\log_{10}$ scale. For comparison, we also provide the results based on the uniform
subsampling. In general, $\pi_i^{\mvc}$ outperforms the uniform
subsampling probability, and its advantage becomes more significant as the tail of
the covariate distribution becomes heavier or if $\tau$ is further
from 0.5.
In general, $\pi_i^{\mvc}$, compared with the uniform subsampling probability, shows a significant
advantage in terms of MSE,
except when $\X$ follows a normal distribution and $\tau=0.5$, 
even though theoretically $\pi_i^{\mvc}$ does not minimize the MSE of the original
parameter.
 We also see that when both the covariate and the response
have heavy tail distributions ($\X$ follows the $t_2$ distribution and
$Y\mid\X$ follows the $t_1$ distribution), the uniform
subsampling probability does not lead to stable results. 

\begin{figure}[htp]
  \centering
\includegraphics[width=0.32\textwidth,page=1]{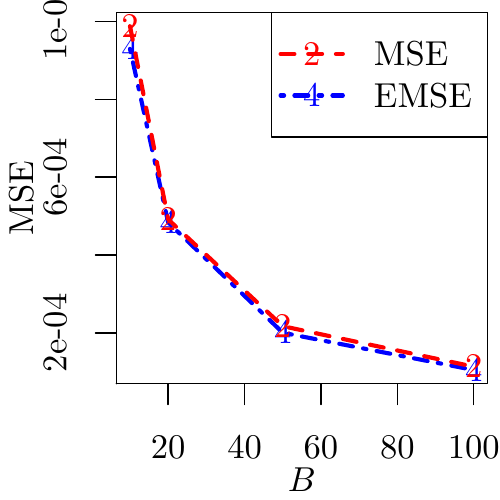}
\includegraphics[width=0.32\textwidth,page=2]{BootEstMse-tau75.pdf}
\includegraphics[width=0.32\textwidth,page=3]{BootEstMse-tau75.pdf}
\includegraphics[width=0.32\textwidth,page=4]{BootEstMse-tau75.pdf}
\includegraphics[width=0.32\textwidth,page=5]{BootEstMse-tau75.pdf}
\includegraphics[width=0.32\textwidth,page=6]{BootEstMse-tau75.pdf}
\includegraphics[width=0.32\textwidth,page=7]{BootEstMse-tau75.pdf}
\includegraphics[width=0.32\textwidth,page=8]{BootEstMse-tau75.pdf}
\includegraphics[width=0.32\textwidth,page=9]{BootEstMse-tau75.pdf}
\caption{Empirical MSE (MSE) and estimated MSE (EMSE) against number
  of repeat subsampling $B$ when $\tau=0.75$. The
  three columns 1-3 correspond to the three distributions of $\X$
  (normal, $t_3$, $t_2$), respectively. The three rows 1-3 correspond to the three conditional distributions of $Y$ (normal, exponential,  $t_1$), respectively.} 
  \label{fig:02}
\end{figure}

To evaluate the performance of the formula in \eqref{eq:23} in
estimating the variance-covariance matrix, we use
$\tr\{\wh\Var(\breve{\bbeta}_I)\}$ to estimate the MSE of
$\breve{\bbeta}_I$, and compare the average estimated MSE with the
empirical MSE. Figure~\ref{fig:02} presents the results for the case
when $\tau=0.75$. 
For all the three different distributions of $\X$ and the three distributions of $Y\mid\X$, the estimated MSEs are very close to the empirical MSEs, indicating that
the proposed formula works well. Results for the case when $\tau=0.5$ are similar and are omitted.

\subsection{Example}

Now we analyze a  data set collected at the ChemoSignals
Laboratory in the BioCircuits Institute, University of California San
Diego. This data set was used to develop and test strategies for
continuously monitoring or improving response time of chemical sensory
systems \citep{fonollosa2015reservoir}.
It contains the readings of 16 chemical sensors exposed to the mixture
of Ethylene and CO at varying concentration levels in the air. Readings
from the second sensor contain about 20\% negative values for
unknown reasons, so we do not use the readings from this sensor.   
For illustration, we model the $\tau=0.75$ quartile  for the readings
from the last sensor using other sensors' readings. As suggested in
\cite{goodson2011mathematical} for chemical concentrations, we take log-transformation of the 
raw data.
The data set was collected over about 12 hours of continuous
operation and we excluded the observations from the first 4 minutes
before the system stabilized.  
Thus, the full data set used contains $N=4,188,261$
observations  with $14$ predictors, and $p=15$ because an intercept is included.

\begin{figure}[htp]
  \centering
\includegraphics[width=0.7\textwidth]{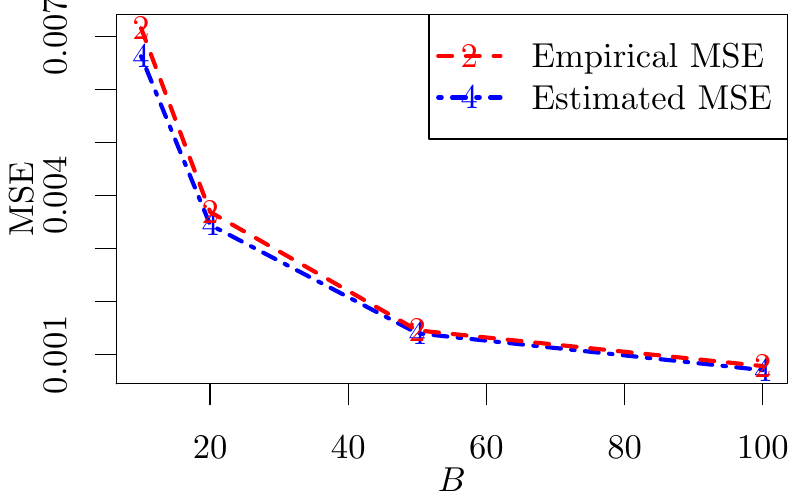}
\caption{Empirical MSE and estimated MSE against number of repeat
  subsampling $B$ for the gas sensor data set. The MSE for the full
  data based on 100 iterations of bootstrapping is
  $1.95\times10^{-5}$.
} 
  \label{fig:03}
\end{figure}

We implement $\breve{\bbeta}_I$ in \eqref{eq:25} with $\pi_i^{\mvc}$, and set
$n_0=1000$, $n=1000$, and $B=10, 20, 50,$ and $ 100$. We repeat the iterative subsampling procedure for $S=1000$ times.  Since the true value of
$\bbeta$ is unknown for a real data set, we use the full data estimate
to access the variation due to subsampling. We calculate the empirical
MSE using
$\textrm{MSE}=S^{-1}\sum_{s=1}^S\|\breve{\bbeta}_I^{(s)}-\wh\bbeta\|^2$,
where $\wh\bbeta=(-0.591,$ $-0.010,$ $-0.725,$ $0.231,$ $-0.433,$
$0.735,$ $0.173,$ $0.554,$ $0.025,$ $-0.009,$ $-0.161,$ $1.052,$
$-0.365,$ $0.048,$ $-0.089)\tp$ for this data set. Figure~\ref{fig:03}
present empirical MSEs and average estimated MSEs for different values
of $B$. The empirical MSE decreases as $B$ increases, indicating
better approximations with larger values of $B$. Furthermore, the
estimated MSEs are very close to the empirical MSEs, showing the
desirable performance of the estimator proposed in \eqref{eq:23}.  

\begin{figure}[htp]
  \centering
\includegraphics[width=\textwidth]{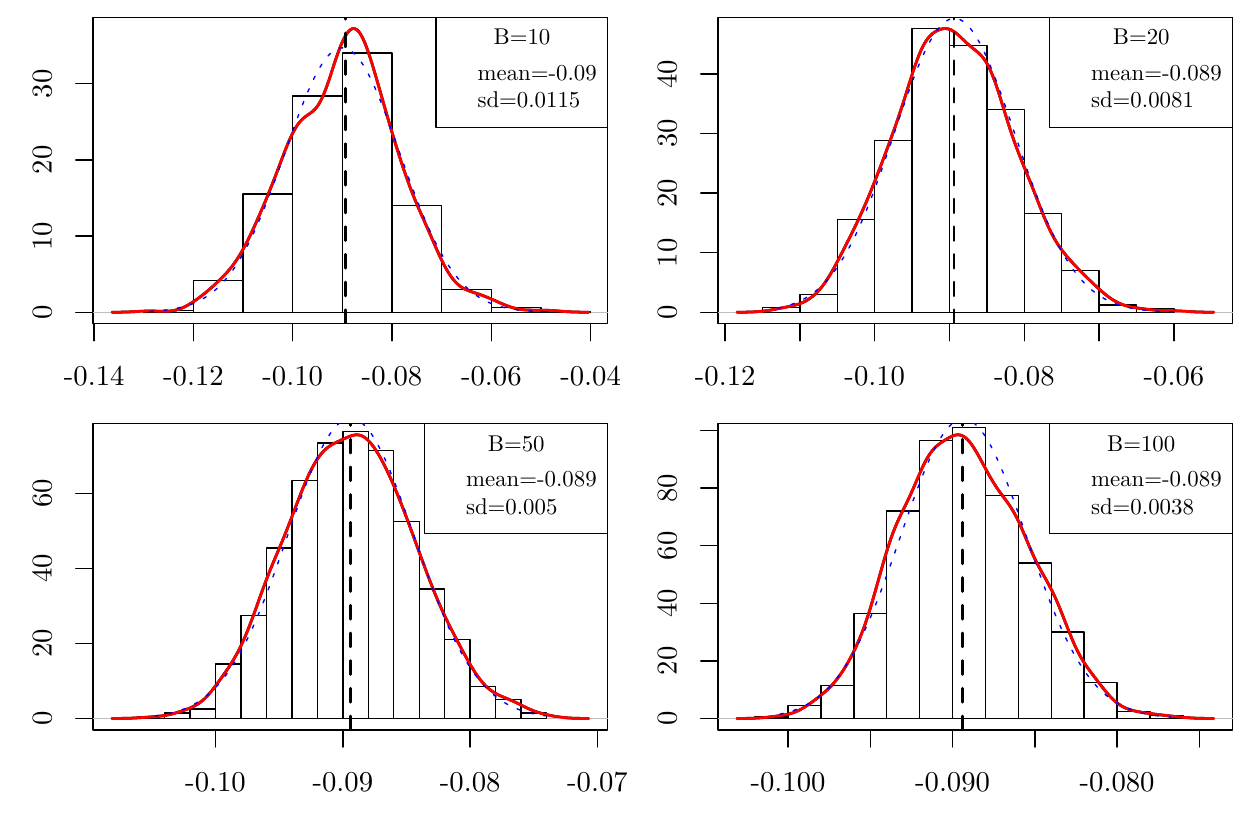}
\caption{Histograms for $\breve{\beta}_{I,14}^{(s)}$s
   with different values of $B$. The vertical dashed line corresponding
  to the value calculated from the full data estimate
  $\wh{\bbeta}_{14}$. The red solid curve is the
  kernel density estimate and the blue dashed curve is the normal
  density curve with the same mean and standard deviation of
  $\breve{\beta}_{I,14}^{(s)}s$.} 
  \label{fig:04}
\end{figure}

To assess the normality of $\breve{\bbeta}_I$, we create histograms
for its last component $\breve{\beta}_{I,14}$.  
Figure~\ref{fig:04} presents results for different values of
$B$. The vertical dashed line corresponds to the value calculated
from the full data estimate, i.e., $\wh{\beta}_{14}$. 
The ``mean'' and ``sd'' in the legend are the mean and
standard deviation for the $S$ values of $\breve{\beta}_{I,14}^{(s)}$.  
The red solid curve is the kernel density estimate based on these $S$
values and the blue dashed curve is the normal density curve with the
same mean and standard deviation. These histograms show clear pattern
of normality, especially for large values of $B$. 

All the calculation were performed on a computer running Ubuntu 18.04
with an Intel I7 CPU. For the full data estimate, using the \verb|rq|
function in the R package \verb|quantreg|, it took the default
algorithm with \verb|br| option over five hours to run. With
  \verb|pfn| option in \verb|br| function, it implements the
  Frisch-Newton approach with preprocessing, in which a pilot estimate
  based on an uniform random subsample is used to preprocess the data 
\citep{portnoy1997gaussian,yang2013quantile}. 
With this method , it took about ten seconds to finish the
calculation. Thus, it is seen that early work on using random
subsampling has already greatly reduced the 
computational burden in quantile regression.
For our Algorithm~\ref{alg:3}, with $n_0=1000$ and $n=1000$, it
took about 0.458 second to approximate the optimal subsampling
probabilities $\pi_i^{\mvc}$. The times used in the second step were
0.65, 1.29, 3.21, and 6.43 seconds for $B=10, 20, 50$, and $100$,
respectively. Thus, the per iteration time cost in Step 2 was about
0.065 second. 
 Note that the Frisch-Newton approach with preprocessing only provides
 a point estimate, whereas Algorithm~\ref{alg:3} also provides
 standard errors for statistical inferences. If we perform estimation
 only, the time to obtain a point estimator is 0.458+0.065, which
 is about 5\% of the time needed for the Frisch-Newton approach with preprocessing.


\section*{Acknowledgement}
The authors are grateful to the editor, associate editor, and two
reviewers for their comments that helped improve the manuscript. The
authors were partially supported by the U.S. National Science
Fundation and the National Institute of Health. 

\section*{Supplementary material}
\label{SM}
Supplementary material available at {\it Biometrika} online includes proofs of all the theoretical results and additional numerical results.

\pagebreak
\begin{center}
\large Supplementary material for\\
\Large  Optimal subsampling for quantile regression in big data\\[2mm]
\large by HaiYing Wang and Yanyuan Ma
\end{center}

In this supplementary material, we prove all the theorems in the main paper and provide additional numerical results.
\setcounter{equation}{0}
\renewcommand{\theequation}{S.\arabic{equation}}
\setcounter{section}{0}
\renewcommand{\thesection}{S.\arabic{section}}
\setcounter{subsection}{0}
\renewcommand{\thesubsection}{S.\arabic{section}-\arabic{subsection}}
\setcounter{figure}{0}
\renewcommand{\thefigure}{S.\arabic{figure}}
\setcounter{table}{0}
\renewcommand{\thetable}{S.\arabic{table}}

\def\boxit#1{\vbox{\hrule\hbox{\vrule\kern6pt\vbox{\kern6pt#1\kern6pt}\kern6pt\vrule}\hrule}}
\def\macomment#1{\vskip 2mm\boxit{\vskip 2mm{\color{red}\bf#1} {\color{blue}\bf -- MA\vskip 2mm}}\vskip 2mm}
\def\wangcomment#1{\vskip 2mm\boxit{\vskip 2mm{\color{blue}\bf#1} {\color{blue}\bf -- Haiying\vskip 2mm}}\vskip 2mm}

\section{Proofs of theorems}
\label{sec:appendix}

\subsection{Proof of Theorem~\ref{thm:1}}
Define
\begin{eqnarray*}
    Z_n^*(\lambda)
  =\sumn\frac{\rho_\tau(\varepsilon_i^*-v_i^*)
  -\rho_\tau(\varepsilon_i^*)}{N\pi_i^*},
\end{eqnarray*}
where $v_i^*=\lambda\tp\x_i^*/\sqrt{n}$ and
$\varepsilon_i^*=y_i^*-\bbeta_t\tp\x_i^*$. As a function of $\lambda$,
$Z_n^*(\lambda)$ is convex and minimized by
$\sqrt{n}(\wt\bbeta-\bbeta_t)$. Thus we can focus on $Z_n^*(\lambda)$
when assessing the properties of $\sqrt{n}(\wt\bbeta-\bbeta_t)$.  

From the following identity 
\begin{eqnarray*}
  \rho_\tau(u-v)-\rho_\tau(u)
  =-v\psi_\tau(u)+\int_0^v\{I(u\le s)-I(u\le 0)\}\ud s,
\end{eqnarray*}
where $\psi_\tau(u)=\tau-I(u<0)$, we have
\be
  Z_n^*(\lambda)
  &=&\sumn\frac{-v_i^*\psi_\tau(\varepsilon_i^*)
     +\int_0^{v_i^*}\{I(\varepsilon_i^*\le s)
     -I(\varepsilon_i^*\le 0)\}\ud s}{N\pi_i^*}\n\\
  &=&\frac{1}{\sqrt{n}}\sumn
     \frac{-\lambda\tp\x_i^*\{\tau-I(\varepsilon_i^*<0)\}}
     {N\pi_i^*}
   +\sumn\frac{\int_0^{v_i^*}\{I(\varepsilon_i^*\le s)
     -I(\varepsilon_i^*\le 0)\}\ud s}{N\pi_i^*}\n\\
  &=&\lambda\tp W_n^*+Z_{2n}^*.\label{eq:3}
\ee
 where
 \begin{eqnarray*}
 W_n^*&=& -\frac{1}{\sqrt{n}}\sumn
      \frac{\x_i^*\{\tau-I(\varepsilon_i^*<0)\}}
      {N\pi_i^*},\\
 Z_{2n}^*&=&\sumn\frac{\int_0^{v_i^*}\{I(\varepsilon_i^*\le s)
      -I(\varepsilon_i^*\le 0)\}\ud s}{N\pi_i^*}.
 \end{eqnarray*}
Denote
\begin{eqnarray*}
  \eta_i^*
  &=&\frac{-\x_i^*\{\tau-I(\varepsilon_i^*<0)\}}{N\pi_i^*},
\end{eqnarray*}
so
$W_n^*=n^{-1/2}\sum_{i=1}^n\eta_i^*$.
We have
\begin{align}
    \Exp(\eta_i^*|\Fn)
  &=\sumN\frac{-\x_i\{\tau-I(\varepsilon_i<0)\}}{N}
    =O_P(1/\sqrt{N}),\label{eq:40}\\
  \Var(\eta_i^*|\Fn)
  &=\sumN\frac{\{\tau-I(\varepsilon_i<0)\}^2\x_i\x_i\tp}{N^2\pi_i}
    -\left[\sumN\frac{-\x_i\{\tau-I(\varepsilon_i<0)\}}{N}\right]^2
  =\V_{\pi} -\op,\label{eq:26}
\end{align}
where $\varepsilon_i=y_i-\bbeta_t\tp\x_i$. In \eqref{eq:40}, $\Exp(\eta_i^*|\Fn)$  is  $O_P(N^{-1/2})
$ because for each element of $\eta_i^*$, say $\eta_{i,j}^*$, 
\begin{align*}
  \Exp\{\Exp(\eta_{i,j}^*|\Fn)\}&=0\\
  \Var\{\Exp(\eta_{i,j}^*|\Fn)\}
  &=\frac{1}{N^2}\sumN\Var\{\tau-I(\varepsilon_i<0)\}x_{i,j}^2
  \le\frac{1}{N^2}\sumN\|\x_i\|^2, 
\end{align*}
and Chebyshev's inequality indicates that $\Exp(\eta_{i,j}^*|\Fn)=O_P(N^{-1/2})$. 

We now check Lindeberg's conditions \citep[Theorem 2.27
of][]{Vaart:98} under the conditional distribution given $\Fn$. 
Specifically,  we want to show that for every $\epsilon>0$,
\begin{align}
    &\sumn\Exp\{\|n^{-1/2}\eta_i^*\|^2
      I(\|\eta_i^*\|>\sqrt{n}\epsilon)\big|\Fn\}\n\\
  &=\sumn
     \Exp\Bigg\{\bigg\|\frac{-\x_i^*\{\tau-I(\varepsilon_i^*<0)\}}
     {\sqrt{n}N\pi_i^*}\bigg\|^2
     I\bigg(\bigg\|\frac{-\x_i^*\{\tau-I(\varepsilon_i^*<0)\}}
     {\sqrt{n}N\epsilon\pi_i^*}\bigg\|>1\bigg)\bigg|\Fn\Bigg\}\n\\
  &=\sumN \frac{\|\x_i\|^2\{\tau-I(\varepsilon_i<0)\}^2} {N^2\pi_i}
     I\bigg(\frac{\|\x_i\||\tau-I(\varepsilon_i<0)|}
     {\sqrt{n}N\epsilon\pi_i}>1\bigg)\n\\
  &\le\sumN \frac{\|\x_i\|^2} {N^2\pi_i}
       I\bigg(\frac{\|\x_i\|}{\sqrt{n}N\epsilon\pi_i}>1\bigg)
       \label{eq:16}
\end{align}
goes to zero  in probability. 
If condition \eqref{eq:17} holds,
then the right hand side of \eqref{eq:16} satisfies that
\be
  \sumN \frac{\|\x_i\|^2} {N^2\pi_i}
    I\bigg(\frac{\|\x_i\|}{\sqrt{n}N\epsilon\pi_i}>1\bigg)
  &\le& \sumN \frac{\|\x_i\|^2} {N^2\pi_i}
    I\bigg(\max_{1\le i\le N}\frac{\|\x_i\|}{\pi_i}
    >\sqrt{n}N\epsilon\bigg)\n\\
  &=&I\bigg(\max_{1\le i\le N}\frac{\|\x_i\|}{\pi_i}
    >\sqrt{n}N\epsilon\bigg) \sumN \frac{\|\x_i\|^2} {N^2\pi_i}
    =\op.\n
\ee
Thus, combining \eqref{eq:26} and Assumption 2 (b), if \eqref{eq:17} 
holds, Lindeberg's conditions hold in probability.

Given $\Fn$, $\eta_i^*$, $i=1,...,n$, are i.i.d with mean
$\Exp(\eta_i^*|\Fn)$ and variance $\Var(\eta_i^*|\Fn)$.
Thus, conditional on $\Fn$, when $n, N\to\infty$,  with probability approaching one, 
\begin{equation*}
  \{\Var(\eta_i^*|\Fn)\}^{-1/2}
  \{W_n^*-\sqrt{n}\Exp(\eta_i^*|\Fn)\}\longrightarrow\Nor(\0, \I), \end{equation*}
in distribution, which implies that
\begin{equation}
  \{\Var(\eta_i^*|\Fn)\}^{-1/2}W_n^*
  \longrightarrow\Nor(\0, \I), \label{eq:44}
\end{equation}
in distribution because $\sqrt{n}\Exp(\eta_i^*|\Fn)\}=O_P(n^{1/2}N^{-1/2})=\op$.

For $Z_{2n}^*$ in \eqref{eq:3}, denote 
$Z_{2ni}^*=\int_0^{v_i^*}\{I(\varepsilon_i^*\le s)-I(\varepsilon_i^*\le
0)\}\ud s$, and
\begin{eqnarray*}
  \Exp(Z_{2ni}^*)
  =\int_0^{v_i^*}\{F_{\varepsilon\mid\X}(s,\x_i^*)-F_{\varepsilon\mid\X}(0,\x_i^*)\}\ud s.
\end{eqnarray*}
The conditional expectation of $Z_{2n}^*$, $\Exp(Z_{2n}^*|\Fn)$, equals
\begin{align}
    &\Exp\left(\sumn\frac{Z_{2ni}^*}{N\pi_i^*}\bigg|\Fn\right)
      = \frac{n}{N}\sumN Z_{2ni}
      =\frac{n}{N}\sumN\Exp(Z_{2ni})
      +\frac{n}{N}\sumN\{Z_{2ni}-\Exp(Z_{2ni})\},\label{eq:5}
  \end{align}
where $Z_{2ni}=\int_0^{v_i}\{I(\varepsilon_i\le s)-I(\varepsilon_i\le
0)\}\ud s$, and
$\Exp(Z_{2ni})=\int_0^{v_i}\{F_{\varepsilon\mid\X}(s,\x_i)-F_{\varepsilon\mid\X}(0,\x_i)\}\ud
s$. For the first term on the right hand side of \eqref{eq:5},
following an approach similar to that in Section 4.2 of
\cite{koenker2005quantile} under the conditions in Assumption 1, we
have 
  \begin{align}
    \frac{n}{N}\sumN\Exp(Z_{2ni})
  =&\frac{n}{N}\sumN\int_0^{v_i}\{F_{\varepsilon\mid\X}(s,\x_i)
  -F_{\varepsilon\mid\X}(0,\x_i)\}\ud s\n\\
  =&\frac{\sqrt{n}}{N}\sumN
    \int_0^{\lambda\tp\x_i}\{F_{\varepsilon\mid\X} (t/{\sqrt{n}},\x_i)-F_{\varepsilon\mid\X}(0,\x_i)\}\ud t\n\\
  =&\frac{1}{N}\sumN\int_0^{\lambda\tp\x_i}
    f_{\varepsilon\mid\X}(0,\x_i)t\ud t + o(1)\n\\
    =&\frac{1}{2N}\sumN(\lambda\tp\x_i)^2f_{\varepsilon\mid\X}(0,\x_i)
       +o(1)\n\\
    =&\frac{1}{2}\lambda\tp D_N\lambda+o(1)
       =\frac{1}{2}\lambda\tp D\lambda+o(1).\label{eq:6}
  \end{align}
The second term in \eqref{eq:5} has mean 0 and variance 
\begin{align}
  \Var\left\{n\sumN\frac{Z_{2ni}-\Exp(Z_{2ni})}{N}\right\}
  &\le\frac{n^2}{N^2}\sumN\Exp(Z_{2ni}^2)\n\\
  & \le\frac{\max_{1\le i\le N}\|\x_i\|}{\sqrt{N}}
   \times \frac{2\|\lambda\|\sqrt{n}}{\sqrt{N}}
    \times\frac{n}{N}\sumN\Exp(Z_{2ni})\label{eq:35}
\end{align}
which, in view of~\eqref{eq:6}, converges to 0 if condition
\eqref{eq:43} holds and $n/N$ does not go to infinity. Here, the
second inequality in (\ref{eq:35}) is from the facts that $Z_{2ni}$ is nonnegative and
  \begin{equation*}
    Z_{2ni}\le\left|\int_0^{\frac{\lambda\tp\x_i}{\sqrt{n}}}
    |\{I(\varepsilon_i\le s)-I(\varepsilon_i\le 0)\}|\ud s\right|
  \le\frac{2|\lambda\tp\x_i|}{\sqrt{n}}.
  \end{equation*}

From \eqref{eq:5}, \eqref{eq:6}, and \eqref{eq:35},
\begin{align}
  \Exp\left\{\sumn\frac{Z_{2ni}^*}{N\pi_i^*}\bigg|\Fn\right\}
  =\frac{n}{N}\sumN Z_{2ni}
  =\frac{1}{2}\lambda\tp D_N\lambda+\op\label{eq:2}
\end{align}
based on Chebyshev's inequality.

Now we exam the conditional variance of $Z_{2n}^*$.  Nothing that conditional on ${\cal F}_N$,
$Z_{2ni}^*$'s are independent and identically distributed, we have 
  \begin{align}
    &\Var\left\{\sumn\frac{Z_{2ni}^*}{N\pi_i^*}\bigg|\Fn\right\}
    \le\frac{n}{N^2} \Exp\left\{\frac{(Z_{2ni}^*)^2}
      {(\pi_i^*)^2}\right\}
    =\frac{n}{N^2}\sumN\frac{Z_{2ni}^2}{\pi_i}\n\\
    &\le\frac{2\sqrt{n}\|\lambda\|}{N^2}\sumN\frac{Z_{2ni}\|\x_i\|}
      {\pi_i}
    \le\max_{1\le i\le N}\frac{\|\x_i\|}{\pi_i}\times
      \frac{2\sqrt{n}\|\lambda\|}{N^2}\sumN Z_{2ni}\n\\
    &=\max_{1\le i\le N}\frac{\|\x_i\|}{\pi_i}\times
      \frac{2\|\lambda\|}{\sqrt{n}N}
      \times\frac{n}{N}\sumN Z_{2ni}.\label{eq:39}
  \end{align}

From \eqref{eq:6}, \eqref{eq:39}, and  condition \eqref{eq:17}, we have
  \begin{align}\label{eq:38}
    \Var\left\{\sumn\frac{Z_{2ni}^*}{N\pi_i^*}\bigg|\Fn\right\}=\op.
  \end{align}

  From (\ref{eq:2}), \eqref{eq:38}, and Chebyshev's inequality,
\begin{equation}
  \sumn\frac{Z_{2ni}^*}{N\pi_i^*}-\frac{1}{2}\lambda\tp D_N\lambda
  =o_{P|\Fn}(1). 
  \label{eq:4}
\end{equation}
Here $a=o_{P|\Fn}(1)$ means $a$ converges to zero in conditional
probability given $\Fn$ in probability, namely, for any $\delta>0$,
$\Pr(|a|>\delta|\Fn)\rightarrow0$ in probability. Note that
$\Pr(|a|>\delta|\Fn)\le1$, thus it converges to 0 in probability if
and only if $\Pr(|a|>\delta) =
\Exp\{\Pr(|a|>\delta|\Fn)\}\rightarrow0$. Therefore, $a=o_{P|\Fn}(1)$
is equivalent to $a=\op$, and we will use the notation of $o_P$ only.

From \eqref{eq:3} and \eqref{eq:4}, we have
\begin{eqnarray*}
  Z_n^*(\lambda)
  =\lambda\tp W_{n}^*+\frac{1}{2}\lambda\tp D_N\lambda+\opf.
\end{eqnarray*} 
Since $Z_n^*(\lambda)$ is convex, from the corollary in page 2 of \cite{hjort2011asymptotics}, its
  minimizer, $\sqrt{n}(\wt\bbeta-\bbeta_t)$, satisfies that
\begin{eqnarray*}
  \sqrt{n}(\wt\bbeta-\bbeta_t)=-D_N^{-1}W_{n}^*+\opf,
\end{eqnarray*}
Thus, we have
\begin{eqnarray*}
  (D_N^{-1}\V_\pi D_N^{-1})^{-1/2}
  \sqrt{n}(\wt\bbeta-\bbeta_t)
  =-(D_N^{-1}\V_\pi D_N^{-1})^{-1/2}D_N^{-1}W_{n}^*+\opf.
\end{eqnarray*}
Combining the the fact that $(D_N^{-1}\V_\pi
D_N^{-1})^{-1/2}D_N^{-1}\V_{\pi}D_N^{-1}(D_N^{-1}\V_\pi
D_N^{-1})^{-1/2}=\I$, the results in~\eqref{eq:26} and \eqref{eq:44},
and Slutsky's Theorem, we have that $(D_N^{-1}\V_\pi D_N^{-1})^{-1/2} 
  \sqrt{n}(\wt\bbeta-\bbeta_t)$ converges to $\Nor(\0,\I)$ in
  conditional distribution given $\Fn$ in probability. This means that
  for any $\x$ 
\begin{align}\label{eq:34}
  \Pr\{(D_N^{-1}\V_\pi D_N^{-1})^{-1/2}\sqrt{n}(\wt\bbeta-\bbeta_t)\le\x|\Fn\}
  \rightarrow \Phi(\x),
\end{align}
in probability, where $\Phi(\x)$ is the cumulative distribution
function of the standard multivariate normal distribution. Note that
the conditional probability in \eqref{eq:34} is a bounded random
variable, thus convergence in probability to a constant implies
convergence in the mean. Therefore, the unconditional probability 
\begin{eqnarray*}
  &&\Pr\{(D_N^{-1}\V_\pi D_N^{-1})^{-1/2}
  \sqrt{n}(\wt\bbeta-\bbeta_t)\le\x\}\\
  &=& \Exp[
  \Pr\{(D_N^{-1}\V_\pi D_N^{-1})^{-1/2}\sqrt{n}(\wt\bbeta-\bbeta_t)\le\x|\Fn\}]
  \rightarrow \Phi(\x).
\end{eqnarray*}
This finishes the proof of Theorem~\ref{thm:1}.

\subsection{Proof of Theorem~\ref{thm:3}}
  Note that
\begin{eqnarray*}
  \tr(\V_\pi )
  &=&\tr\left[
    \sumN\frac{\{\tau-I(\varepsilon_i<0)\}^2\x_i\x_i\tp}{N^2\pi_i}
    \right]\\
  &=&\frac{1}{N^2}\sumN\tr\left[
    \frac{\{\tau-I(\varepsilon_i<0)\}^2\x_i\x_i\tp }{\pi_i}
    \right]\\
  &=&\frac{1}{N^2}\sumN
    \frac{\{\tau-I(\varepsilon_i<0)\}^2\|\x_i\|^2}{\pi_i}\\
  &=&\frac{1}{N^2}\left(\sumN\pi_i\right)\left(\sumN
    \frac{\{\tau-I(\varepsilon_i<0)\}^2\|\x_i\|^2}{\pi_i}\right)\\
  &\ge&\frac{1}{N^2}\left\{\sumN
    |\tau-I(\varepsilon_i<0)|\|\x_i\| \right\}^2,
\end{eqnarray*}
 where the last step is from the Cauchy-Schwarz inequality and the
 equality in it holds if and only if when
 $\pi_i\propto|\tau-I(\varepsilon_i<0)|\|\x_i\|$. 

\subsection{Proof of Theorem~\ref{thm:2}}
  Note that
\begin{eqnarray*}
  \tr(D_N^{-1}\V_\pi D_N^{-1})
  &=&\tr\left[D_N^{-1}
    \sumN\frac{\{\tau-I(\varepsilon_i<0)\}^2\x_i\x_i\tp}{N^2\pi_i}
    D_N^{-1}\right]\\
  &=&\frac{1}{N^2}\sumN\tr\left[
    \frac{\{\tau-I(\varepsilon_i<0)\}^2D_N^{-1}\x_i\x_i\tp D_N^{-1}}{\pi_i}
    \right]\\
  &=&\frac{1}{N^2}\sumN
    \frac{\{\tau-I(\varepsilon_i<0)\}^2\|D_N^{-1}\x_i\|^2}{\pi_i}\\
  &=&\frac{1}{N^2}\left(\sumN\pi_i\right)\left(\sumN
    \frac{\{\tau-I(\varepsilon_i<0)\}^2\|D_N^{-1}\x_i\|^2}{\pi_i}\right)\\
  &\ge&\frac{1}{N^2}\left\{\sumN
    |\tau-I(\varepsilon_i<0)|\|D_N^{-1}\x_i\| \right\}^2,
\end{eqnarray*}
 where the last step is from the Cauchy-Schwarz inequality and the
 equality in it holds if and only if when
 $\pi_i\propto|\tau-I(\varepsilon_i<0)|\|D_N^{-1}\x_i\|$. 

\subsection{Proof of Theorem~\ref{thm:4}}
Recall the notations $v_i^*=\lambda\tp\x_i^*/\sqrt{n}$,
$\varepsilon_i^*=y_i^*-\bbeta_t\tp\x_i^*$, and
$\psi_\tau(u)=\tau-I(u<0)$. Let
\begin{align}
  \breve{Z}_n(\lambda)
  =&\sumn\frac{\rho_\tau(\varepsilon_i^*-v_i^*)
  -\rho_\tau(\varepsilon_i^*)}{N\pi_i^{*\wt\bbeta_0}},\n\\
  =&\sumn\frac{-v_i^*\psi_\tau(\varepsilon_i^*)
     +\int_0^{v_i^*}\{I(\varepsilon_i^*\le s)
     -I(\varepsilon_i^*\le 0)\}\ud s}{N\pi_i^{*\wt\bbeta_0}},\n\\
  =&\frac{1}{\sqrt{n}}\sumn
     \frac{-\lambda\tp\x_i^*\{\tau-I(\varepsilon_i^*<0)\}}
     {N\pi_i^{*\wt\bbeta_0}}
   +\sumn\frac{\int_0^{v_i^*}\{I(\varepsilon_i^*\le s)
     -I(\varepsilon_i^*\le 0)\}\ud s}{N\pi_i^{*\wt\bbeta_0}},\n\\
  \equiv&\lambda\tp\breve{W}_{n}^*+\breve{Z}_{2n}^*.\label{eq:10}
\end{align}
Denote
\begin{eqnarray*}
  \breve\eta_i^*
  =\frac{-\x_i^*\{\tau-I(\varepsilon_i^*<0)\}}{N\pi_i^{*\wt\bbeta_0}}.
\end{eqnarray*} 
We have
\begin{align}
  \Exp(\breve\eta_i^*|\Fn,\wt\bbeta_0)
  =&\sumN\frac{-\x_i\{\tau-I(\varepsilon_i<0)\}}{N}= O_P(N^{-1/2})\n\\
  \Var(\breve\eta_i^*|\Fn,\wt\bbeta_0)
  =&\sumN\frac{\{\tau-I(\varepsilon_i<0)\}^2\x_i\x_i\tp}
     {N^2\pi_i^{\wt\bbeta_0}}-\op\label{eq:31}
\end{align}

For $\pi_i^{\mmse}$,  $\pi_i^{\wt\bbeta_0}=\pi_{i}^{\mmse}(\wt\bbeta_0)$, and we have
\begin{align}
 &\sumN\frac{\{\tau-I(\varepsilon_i<0)\}^2\x_i\x_i\tp}
     {N^2\pi_{i}^{\mmse}(\wt\bbeta_0)}\n\\
  =&\oneN\sumN\frac{\{\tau-I(\varepsilon_i<0)\}^2\x_i\x_i\tp}
     {|\tau-I(\varepsilon_i^{\wt\bbeta_0}<0)|
     \|\wt{D}_N^{-1}\x_i\|}\times
     \oneN\sumN|\tau-I(\varepsilon_i^{\wt\bbeta_0}<0)|
     \|\wt{D}_N^{-1}\x_i\|\equiv\wt\Delta_1\times\wt\Delta_2.
     \label{eq:27}
\end{align}
Now we show that $\wt\Delta_1-\Delta_1=\op$ and
$\wt\Delta_2-\Delta_2=\op$, where $\Delta_1$ and $\Delta_2$ have the
same expression of $\wt\Delta_1$ and $\wt\Delta_2$, respectively, 
except that $\varepsilon_i^{\wt\bbeta_0}=y_i-\wt\bbeta_0\tp\x_i$ and
$\wt{D}_N$ are replaced by $\varepsilon_i$ and $D_N$,
respectively. Denote $\tau_m=\min(\tau,1-\tau)$. For the $j_1,j_2$th
element of $\wt\Delta_1-\Delta_1$, $j_1,j_2=1,...,p$,  
\begin{align}
  |\wt\Delta_1-\Delta_1|_{j_1,j_2}\le
  &\oneN\sumN\Bigg|
    \frac{\{\tau-I(\varepsilon_i<0)\}^2\|\x_i\|^2}
     {|\tau-I(\varepsilon_i^{\wt\bbeta_0}<0)|\|{D}_N^{-1}\x_i\|}
    -\frac{\{\tau-I(\varepsilon_i<0)\}^2\|\x_i\|^2}
     {|\tau-I(\varepsilon_i<0)|\|D_N^{-1}\x_i\|}\Bigg|\n\\
  &+\oneN\sumN\Bigg|\frac{\{\tau-I(\varepsilon_i<0)\}^2\|\x_i\|^2}
     {|\tau-I(\varepsilon_i^{\wt\bbeta_0}<0)|
    \|\wt{D}_N^{-1}\x_i\|}
    -\frac{\{\tau-I(\varepsilon_i<0)\}^2\|\x_i\|^2}
     {|\tau-I(\varepsilon_i^{\wt\bbeta_0}<0)|
     \|D_N^{-1}\x_i\|}\Bigg|\label{eq:28}\\
  \le&\oneN\sumN\Bigg|\frac{|\tau-I(\varepsilon_i<0)|\|\x_i\|^2}
     {|\tau-I(\varepsilon_i^{\wt\bbeta_0}<0)|
    \|D_N^{-1}\x_i\|}
    -\frac{|\tau-I(\varepsilon_i^{\wt\bbeta_0}<0)|\|\x_i\|^2}
     {|\tau-I(\varepsilon_i^{\wt\bbeta_0}<0)|
     \|D_N^{-1}\x_i\|}\Bigg|+\op\n\\
  \le&\frac{1}{\tau_mN}\sumN
       \frac{|I(\varepsilon_i^{\wt\bbeta_0}<0)-I(\varepsilon_i<0)|\|\x_i\|^2}
     {\|D_N^{-1}\x_i\|}+\op\n\\
  \le&\frac{\lambda_{\max}^D\{1+\op\}}{\tau_mN}\sumN
       |I(\varepsilon_i^{\wt\bbeta_0}<0)-I(\varepsilon_i<0)|\|\x_i\|
       +\op,\label{eq:29}
\end{align}
in which $\lambda_{\max}^D$ is the largest eigenvalue of $D$. 
Here, the term in \eqref{eq:28} is $\op$ due to the uniform
convergence of $\wt{f}_{\varepsilon\mid\X}(0,\x)$; 
 the second last inequality also used $\wt\bb_0-\bb_t=o_p(1)$; and
 $D_N$ can be replaced
in the last step by its limit $D$ because of condition
\eqref{eq:42}. For any $\epsilon>0$, 
\begin{align}
  &\Pr\Bigg\{\oneN\sumN
    |I(\varepsilon_i^{\wt\bbeta_0}<0)-I(\varepsilon_i<0)|\|\x_i\|
    >\epsilon\Bigg\}\n\\
  &\le\frac{1}{\epsilon N}\sumN
       \Exp\{|I(\varepsilon_i^{\wt\bbeta_0}<0)
       -I(\varepsilon_i<0)|\}\|\x_i\|.\label{eq:30}
\end{align}
Note that for each $i$, $|I(\varepsilon_i^{\wt\bbeta_0}<0)
       -I(\varepsilon_i<0)|$ is bounded and converges in probability
       to 0, as $n_0\rightarrow\infty$ and $n\rightarrow\infty$. Thus,
       $\Exp\{|I(\varepsilon_i^{\wt\bbeta_0}<0) 
       -I(\varepsilon_i<0)|\}\rightarrow0$. This indicates that the
       term in \eqref{eq:30} converges to 0, which implies that the
       term in \eqref{eq:29} converges in probability to 0. Thus,
       $\wt\Delta_1-\Delta_1=\op$. Using a similar approach, it can be
       shown that $\wt\Delta_2-\Delta_2=\op$. These facts, together
       with \eqref{eq:31} and \eqref{eq:27}, show that, for
       $\pi_i^{\mmse}$, 
\begin{align}
  \Var(\breve\eta_i^*|\Fn,\wt\bbeta_0)
  =&\sumN\frac{\{\tau-I(\varepsilon_i<0)\}^2\x_i\x_i\tp}
     {N^2\pi_i^{\mmse}}+\opfb.
\end{align}

For $\pi_i^{\mvc}$,  $\pi_i^{\wt\bbeta_0}=\pi_{i}^{\mvc}(\wt\bbeta_0)$, and we have
\begin{align}
  &\sumN\frac{\{\tau-I(\varepsilon_i<0)\}^2\x_i\x_i\tp}
     {N^2\pi_{i}^{\mvc}(\wt\bbeta_0)}\n\\
  &=\oneN\sumN\frac{\{\tau-I(\varepsilon_i<0)\}^2\x_i\x_i\tp}
     {|\tau-I(\varepsilon_i^{\wt\bbeta_0}<0)|\|\x_i\|}\times
     \oneN\sumN|\tau-I(\varepsilon_i^{\wt\bbeta_0}<0)|
    \|\x_i\|\equiv\wt\Delta_3\times\wt\Delta_4.
    \label{eq:32}
\end{align}
Now we show that $\wt\Delta_3-\Delta_3=\op$ and
$\wt\Delta_4-\Delta_4=\op$, where $\Delta_3$ and $\Delta_4$ have the
same expression of $\wt\Delta_3$ and $\wt\Delta_4$, respectively,
except that $\varepsilon_i^{\wt\bbeta_0}$ is replaced by
$\varepsilon_i$. Note that the $j_1,j_2$th element of
$\wt\Delta_3-\Delta_3$ or $\wt\Delta_4-\Delta_4$, $j_1,j_2=1,...,p$,
is bounded by 
\begin{align}\label{eq:45}
  \frac{1}{\tau_mN}\sumN
  |I(\varepsilon_i^{\wt\bbeta_0}<0)-I(\varepsilon_i<0)|\|\x_i\|.
\end{align}
 Using a similar approach used for the case of $\pi_i^{\mmse}$,
 \eqref{eq:45} can be shown to be $\opfb$.  Thus,
 $\wt\Delta_3-\Delta_3=\op$ and $\wt\Delta_4-\Delta_4=\op$. These
 facts, together with \eqref{eq:31}, yield that, for $\pi_i^{\mvc}$, 
\begin{align}
  \Var(\breve\eta_i^*|\Fn,\wt\bbeta_0)
  =&\sumN\frac{\{\tau-I(\varepsilon_i<0)\}^2\x_i\x_i\tp}
     {N^2\pi_i^{\mvc}}+\opfb.
\end{align}

We now check the Lindeberg's condition given $\Fn$ and
  $\wt\bbeta_0$.\\ 
For every $\epsilon>0$,
\begin{align}
  &\sumn\Exp(\|n^{-1/2}\breve\eta_i^*\|^2
    I(\|\breve\eta_i^*\|>\sqrt{n}\epsilon)\big|\Fn,\wt\bbeta_0)\n\\
  =&\sumn
     \Exp\Bigg[\bigg\|\frac{-\x_i^*\{\tau-I(\varepsilon_i^*<0)\}}
     {\sqrt{n}N\pi_i^{*\wt\bbeta_0}}\bigg\|^2
     I\bigg(\bigg\|\frac{-\x_i^*\{\tau-I(\varepsilon_i^*<0)\}}
     {\sqrt{n}N\epsilon\pi_i^{*\wt\bbeta_0}}\bigg\|>1\bigg)
     \bigg|\Fn,\wt\bbeta_0\Bigg]\n\\
  =&\sumN \frac{\|\x_i\|^2\{\tau-I(\varepsilon_i<0)\}^2}
     {N^2\pi_i^{\wt\bbeta_0}}
     I\bigg(\frac{\|\x_i\||\tau-I(\varepsilon_i<0)|}
     {\sqrt{n}N\epsilon\pi_i^{\wt\bbeta_0}}>1\bigg)\n\\
  \le&\frac{1}{N^2}\sumN\frac{\|\x_i\|^2}{\pi_i^{\wt\bbeta_0}}
       I\bigg(\frac{\|\x_i\|}
       {\sqrt{n}N\epsilon\pi_i^{\wt\bbeta_0}}>1\bigg)
       \le I\bigg(\max_{1\le i\le N}\frac{\|\x_i\|}
       {\sqrt{n}N\epsilon\pi_i^{\wt\bbeta_0}}>1\bigg)
       \frac{1}{N^2}
       \sumN\frac{\|\x_i\|^2}{\pi_i^{\wt\bbeta_0}}.\label{eq:18}
\end{align}
Now we show that the term on the right-hand-size of \eqref{eq:18} is $\op$. 

For $\pi_i^{\mmse}$,
\begin{align*}
  \frac{\|\x_i\|}{\pi_i^{\wt\bbeta_0}}
  =& \frac{\|\x_i\|}{|\tau-I(\varepsilon_i^{\wt\bbeta_0}<0)|
     \|\wt{D}_N^{-1}\x_i\|}
     \sum_{j=1}^{N}|\tau-I(\varepsilon_j^{\wt\bbeta_0}<0)|
     \|\wt{D}_N^{-1}\x_j\|\\
  \le&\frac{N}{\tau_m}\frac{\|\x_i\|}{\|\wt{D}_N^{-1}\x_i\|}
       \times\oneN\sum_{j=1}^{N}\|\wt{D}_N^{-1}\x_j\|\\
  \le&\frac{N\lambda_{\max}^{\wt{D}_N}}
       {\tau_m\lambda_{\min}^{\wt{D}_N}}
       \times\frac{\|\x_i\|}{\|\x_i\|}
       \times\oneN\sum_{j=1}^{N}\|\x_j\|
  \le\frac{N\lambda_{\max}^{D}\sqrt{\tr(D_0)}}
       {\tau_m\lambda_{\min}^{D}}\{1+\op\},
\end{align*}
where the $\op$ does not depend on $i$. Thus, 
\begin{equation}
  \max_{1\le i\le N}\frac{\|\x_i\|}{\pi_i^{\wt\bbeta_0}}
\le\frac{N\lambda_{\max}^{D}\sqrt{\tr(D_0)}}
       {\tau_m\lambda_{\min}^{D}}\{1+\op\}.
  \label{eq:33}
\end{equation}
From \eqref{eq:18} and \eqref{eq:33}, 
\begin{eqnarray*}
  &&\sumn\Exp(\|n^{-1/2}\breve\eta_i^*\|^2
    I(\|\breve\eta_i^*\|>\sqrt{n}\epsilon)\big|\Fn,\wt\bbeta_0)\n\\
  &\le&\frac{1}{N^2}\sumN\frac{\|\x_i\|^2}{\pi_i^{\wt\bbeta_0}}
       I\bigg(\max_{1\le i\le N}\frac{\|\x_i\|}
       {\sqrt{n}N\epsilon\pi_i^{\wt\bbeta_0}}>1\bigg)\n\\
  &\le &I\bigg(\frac{\lambda_{\max}^{D}\sqrt{\tr(D_0)}}
       {\sqrt{n}\epsilon\tau_m\lambda_{\min}^{D}}\{1+\op\}>1\bigg)
       \frac{1}{N^2}\sumN\frac{\|\x_i\|^2}{\pi_i^{\wt\bbeta_0}}=\op.
\end{eqnarray*}

For $\pi_i^{\mvc}$, 
\begin{align}\label{eq:48}
  \frac{\|\x_i\|}{\pi_i^{\wt\bbeta_0}}
  =\frac{\|\x_i\|}{|\tau-I(\varepsilon_i^{\wt\bbeta_0}<0)|\|\x_i\|}
     \sum_{j=1}^{N}|\tau-I(\varepsilon_j^{\wt\bbeta_0}<0)|\|\x_j\|
  \le\frac{1}{\tau_m}\sum_{j=1}^{N}\|\x_j\|.
\end{align}
Thus, using an approach similar to that used for the case of
$\pi_i^{\mmse}$, the right hand side of \eqref{eq:18} is $\op$.

Given $\Fn$ and $\wt\bbeta_0$, $\breve\eta_i^*$, $i=1,...,n$, are
i.i.d with mean $\op$ and variance $\V_{\opt}+\op$, where $\V_{\opt}$
has the expression of $\V_{\mvc}$ in \eqref{eq:46} for $\pi_i^{\mvc}$
or $\V_{\mvc}$ in \eqref{eq:47} for $\pi_i^{\mmse}$. Note that
  if $N^{-1}\sumN\|\x_i\|^{-1}{\x_i\x_i\tp}$ is asymptotically
  positive definite, then $\V_{\opt}$ is asymptotically positive
  definite because $|\tau-I(\varepsilon_i<0)|$ is bounded away from
  both 0 and infinity and $D_N$ converges to a finite positive definite matrix.
Thus, given $\Fn$ and $\wt\bbeta_0$ in probability, as
$n_0\rightarrow\infty$, $n\rightarrow\infty$, and
$N\rightarrow\infty$, if $n/N\rightarrow0$, then
\begin{eqnarray*}
  \V_{\opt}^{-1/2}\breve{W}_{n}^*\longrightarrow \Nor(\0, \I),
\end{eqnarray*}
in distribution.

Note that
$Z_{2ni}^*=\int_0^{v_i^*}\{I(\varepsilon_i^*\le s)-I(\varepsilon_i^*\le
0)\}\ud s$ and
$\Exp(Z_{2ni})^*=\int_0^{v_i^*}\{F_i^*(s)-f_{\varepsilon\mid\X}(0,\x_i^*)\}\ud
s$. For the second term in \eqref{eq:10}, i.e. $\breve{Z}_{2n}^*$, we have
\begin{align}
  \Exp(\breve{Z}_{2n}^*|\Fn,\wt\bbeta_0)=
  &\Exp\left\{\sumn\frac{Z_{2ni}^*}{N\pi_i^{*\wt\bbeta_0}}
    \bigg|\Fn,\wt\bbeta_0\right\}
    = \frac{n}{N}\sumN Z_{2ni}
    =\frac{1}{2}\lambda\tp D_N\lambda+\op,\label{eq:49}
  \end{align}
where the last equality is from \eqref{eq:2}. 

Now we exam its variance, which is
  \begin{eqnarray*}
    &&\Var\left\{\sumn\frac{Z_{2ni}^*}{N\pi_i^{*\wt\bbeta_0}}
    \bigg|\Fn,\wt\bbeta_0\right\}
    \le\frac{n}{N^2}\Exp\left\{\frac{(Z_{2ni}^*)^2}
      {(\pi_i^{*\wt\bbeta_0})^2}\bigg|\Fn,\wt\bbeta_0\right\}
    =\frac{n}{N^2}\sumN\frac{Z_{2ni}^2}{\pi_i^{\wt\bbeta_0}}\n\\
    &\le&\frac{2\sqrt{n}\|\lambda\|}{N^2}\sumN\frac{Z_{2ni}\|\x_i\|}
      {\pi_i^{\wt\bbeta_0}}
    \le\max_{1\le i\le N}\frac{\|\x_i\|}{\pi_i^{\wt\bbeta_0}}\times
      \frac{2\sqrt{n}\|\lambda\|}{N^2}\sumN Z_{2ni}\n\\
    &=&\max_{1\le i\le N}\frac{\|\x_i\|}{\pi_i^{\wt\bbeta_0}}\times
      \frac{2\|\lambda\|}{\sqrt{n}N}
      \times\frac{n}{N}\sumN Z_{2ni}.
  \end{eqnarray*}

Considering \eqref{eq:33} or \eqref{eq:48}, corresponding to
$\pi_i^{\mmse}$ or $\pi_i^{\mvc}$, respectively, and results in
\eqref{eq:2},  
we have
  \begin{align}
    \Var(\breve{Z}_{2n}^*|\Fn,\wt\bbeta_0)
    =\Var\left\{\sumn\frac{Z_{2ni}^*}{N\pi_i^{*\wt\bbeta_0}}
    \bigg|\Fn,\wt\bbeta_0\right\}=O_P(n^{-1/2}).\label{eq:50}
  \end{align}

  From (\ref{eq:49}), \eqref{eq:50}, and Chebyshev's inequality,
\begin{equation}
  \sumn\frac{Z_{2ni}^*}{N\pi_i^{*\wt\bbeta_0}}-\frac{1}{2}\lambda\tp D_N\lambda=\opfb.
  \label{eq:51}
\end{equation}
From \eqref{eq:10} and \eqref{eq:51},
\begin{equation}\label{eq:20}
    \breve{Z}_n^*(\lambda)
  =\lambda\tp\breve{W}_{n}^*+\frac{1}{2}\lambda\tp D_N\lambda+\opfb.
\end{equation}

Since $\breve{Z}_n^*(\lambda)$ is convex, from the corollary in page 2
of \cite{hjort2011asymptotics}, its minimizer,
$\sqrt{n}(\breve\bbeta-\bbeta_t)$, satisfies that 
\begin{equation*}
  \sqrt{n}(\breve\bbeta_{\opt}-\bbeta_t)=-D_N^{-1}\breve{W}_{n}^*+\op,
\end{equation*}
where $\breve\bbeta_{\opt}=\breve\bbeta_{\mvc}$ for $\pi_i^{\mvc}$ and
$\breve\bbeta_{\opt}=\breve\bbeta_{\mmse}$ for $\pi_i^{\mmse}$.
Thus, we have
\begin{align*}
  (D_N^{-1}\V_{\opt} D_N^{-1})^{-1/2}
  \sqrt{n}(\breve\bbeta_{\opt}-\bbeta_t)
  =-(D_N^{-1}\V_{\opt} D_N^{-1})^{-1/2}D_N^{-1}\breve{W}_{n}^*+\opf,
\end{align*}
which implies that $(D_N^{-1}\V_{\opt} D_N^{-1})^{-1/2}
  \sqrt{n}(\breve\bbeta_{\opt}-\bbeta_t)$ converges to $\Nor(\0,\I)$ in
  conditional distribution given $\Fn$ and $\wt\bbeta_0$ in
  probability, meaning that for any $\x$ 
\begin{align*}
  \Pr\{(D_N^{-1}\V_{\opt} D_N^{-1})^{-1/2}
  \sqrt{n}(\breve\bbeta_{\opt}-\bbeta_t)\le\x|\Fn,\wt\bbeta_0\}
  \rightarrow \Phi(\x),
\end{align*}
in probability. Since the conditional probability is a bounded random
variable, convergence in probability to a constant implies convergence
in the mean. Therefore, the unconditional probability converges and
this finishes the proof of Theorem~\ref{thm:4}.
\newpage

\section{Additional numerical results}
\subsection{Multiple quantile levels, including some extreme levels}
\label{sec:mult-quant-levels}
In this section, we carry out simulations to assess the performance of
the proposed method in comparison with the full data estimator at
multiple quantile levels, including some extreme levels. Specifically,
we let $\tau=0.01, 0.02, 0.05, 0.1, 0.3, 0.5, 0.7, 0.9, 0.95, 0.98$
and $0.99$. We set the full data sample size $N=10^6$; set the pilot
subsample size $n_0=10^3$; and set the subsample size $n=10^3, 2\times
10^3, 3\times 10^3$, and $5\times 10^3$ with $B=10$ so the total
subsample sizes are $n=10^4, 2\times 10^4, 3\times 10^4$, and $5\times
10^4$, respectively. The same model setup as presented in Section
\ref{sec:numer-exper} of the main paper is used here.  

\begin{figure}[H]
  \centering
\includegraphics[width=0.9\textwidth,page=1]{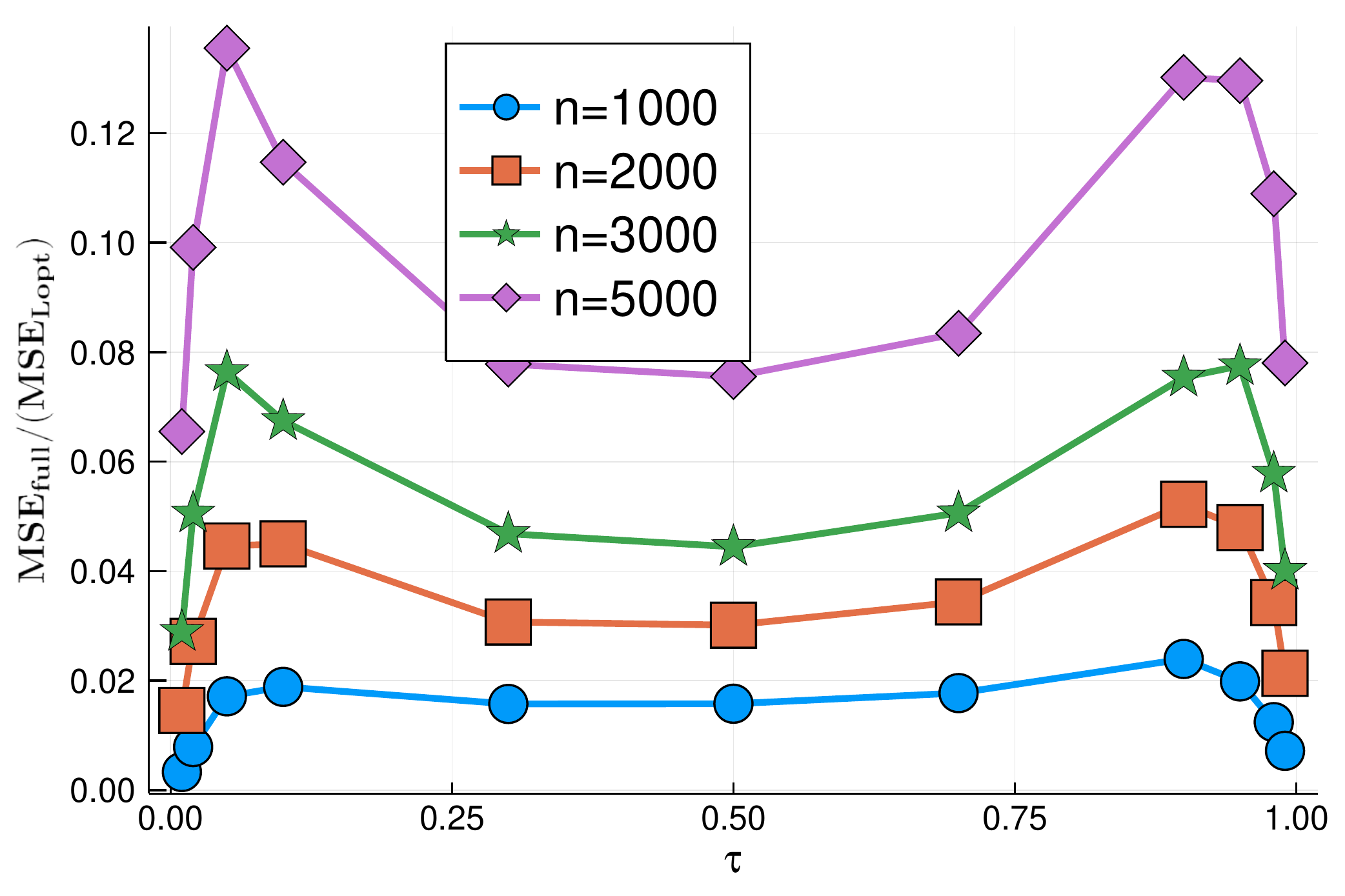}\\
\includegraphics[width=0.9\textwidth,page=1]{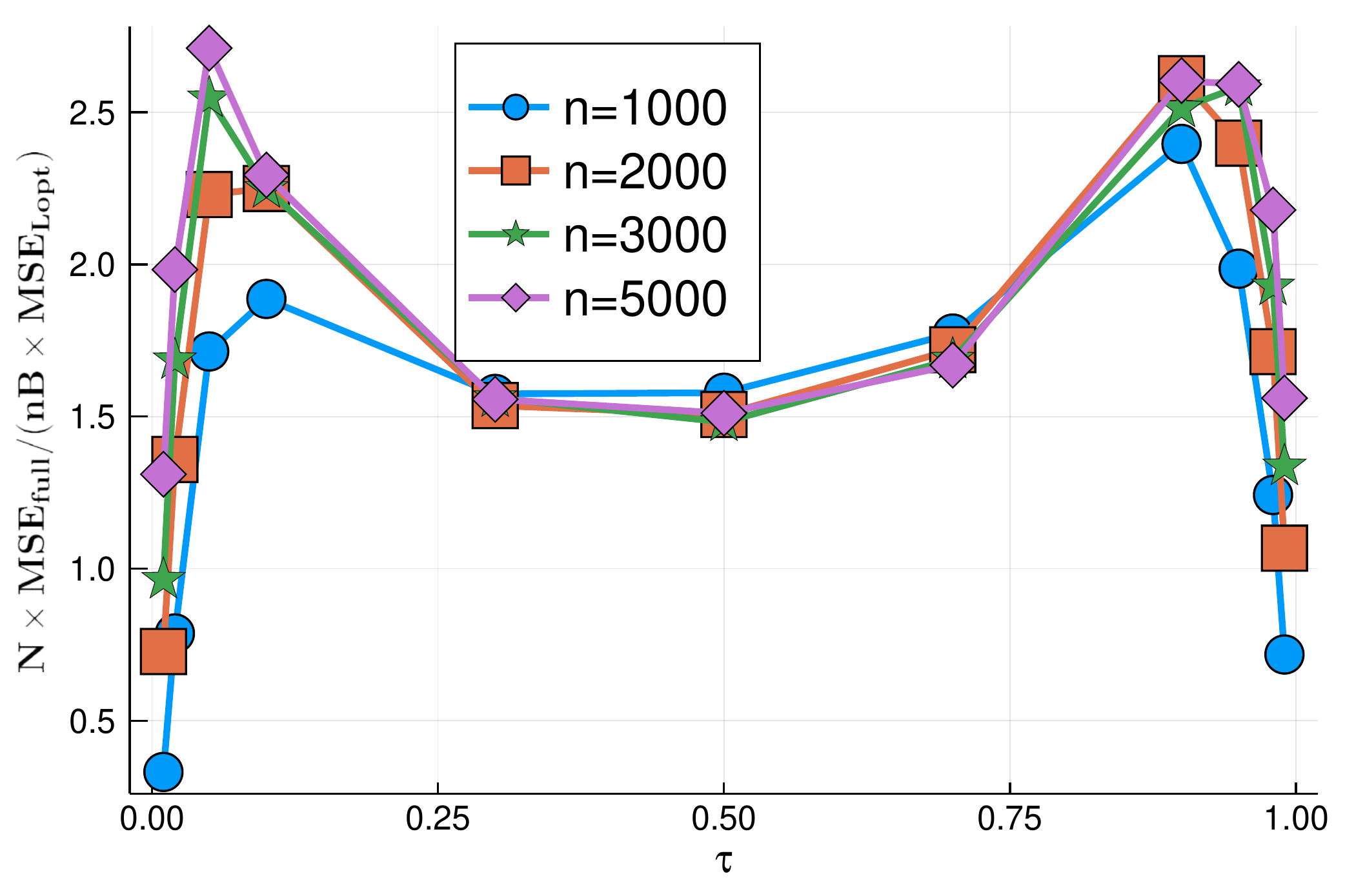}
\caption{Ratio of MSE for the subsample estimator to that of the full
  data estimator against $\tau$. Here $B=10$ and $n$ are set to
  four different values, $\X\sim t_3$, and the conditional
  distributions of $Y$ is exponential.}  
  \label{fig:s1}
\end{figure}

To evaluate the relative efficiency of the proposed method compared
with the full data estimator, the first plot in Figure \ref{fig:s1}
presents the relative performance
${\text{MSE}_{\text{full}}}/{\text{MSE}_{\text{Lopt}}}$. 
Clearly, as the subsample size $n$ increases, the estimation
efficiency of the proposed method gets higher.  
It is also seen that all relative MSEs are smaller than one, meaning
that the performance in terms of MSE of the full data estimator is
always better than that of the subsample estimator, regardless of the
quantile level. This is 
because the subsample based analysis provides
estimators at $\sqrt{nB}$-rate while the full data based analysis
generates estimators at $\sqrt{N}$-rate. 

To eliminate the effect from different sample sizes,  we also reported
the sample size adjusted MSE ratio, $(N\
\text{MSE}_{\text{full}})/(nB\ \text{MSE}_{\text{Lopt}})$, 
in the second plot of Figure \ref{fig:s1} for more informative comparisons.  This ratio can
be interpreted as a measure to compare the per-observation efficiency
between the proposed method and the full data analysis. 
It is seen that most of the ratios are larger than one, expect for
extreme quantile levels such as $\tau=0.01$ and $\tau=0.99$. This
indicates that 
smaller sample size is hardly enough to provide useful information for
extreme quantile levels. As soon as the sample size is sufficient to
perform meaningful analysis, the adjusted MSE improves very
fast. Interestingly, as soon as 
the sample size is reasonably large for the corresponding quantile
estimation, the subsample analysis tends to outperform the full data
analysis in terms of adjusted MSE. This is because the optimized
subsampling probabilities select better subsamples for which the
observations are on average more informative than the observations in
the full data.

\subsection{Sensitivity with respect to the subsample size}\label{sec:sensup}

In this section, we investigate the sensitivity issue of the proposed
method with respect to subsample size $n$. We consider two scenarios:
one with relatively large subsample sizes and one with small subsample
sizes. 

\begin{figure}[H]
  \centering
\includegraphics[width=0.485\textwidth]{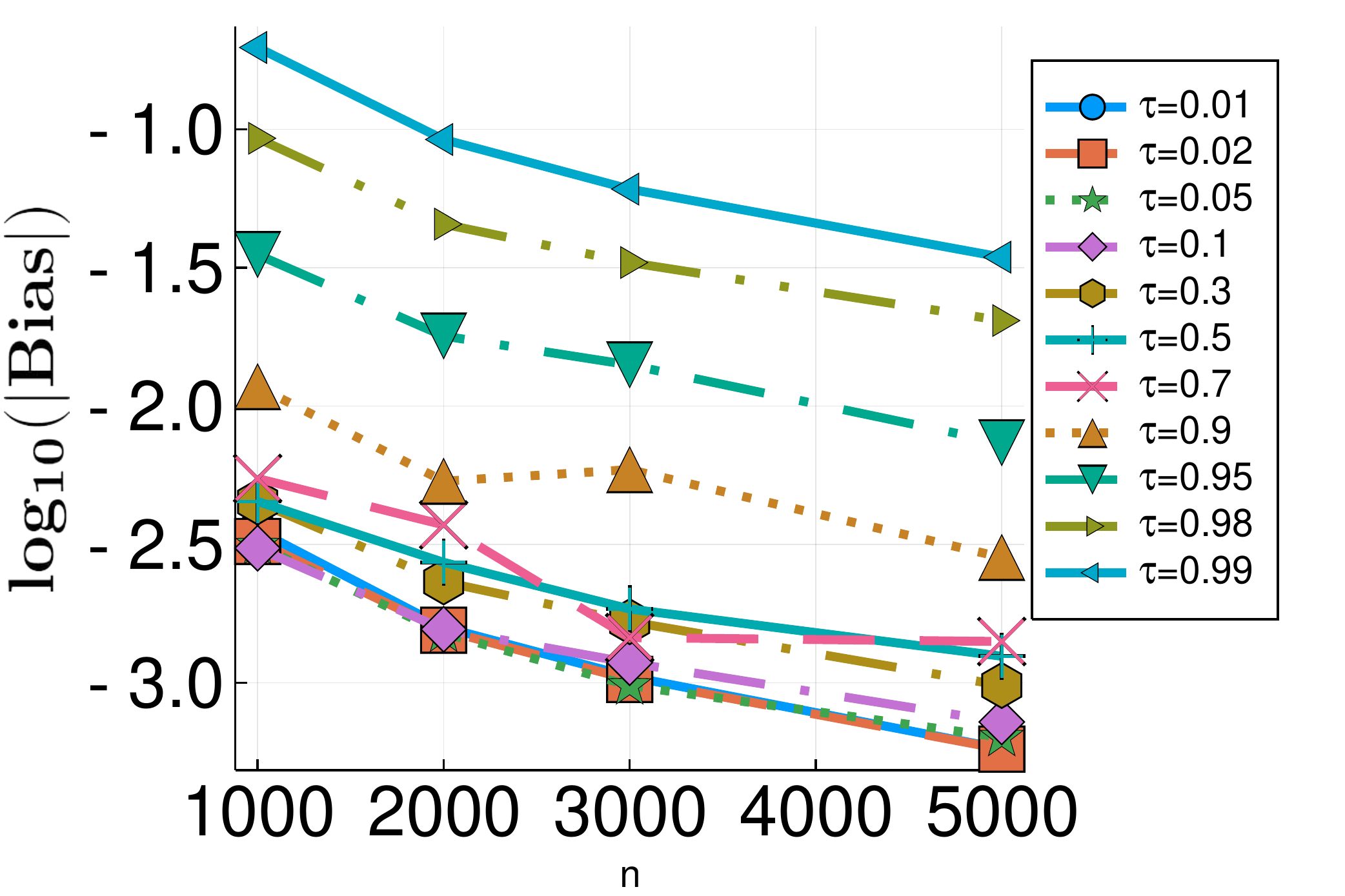}
\includegraphics[width=0.485\textwidth]{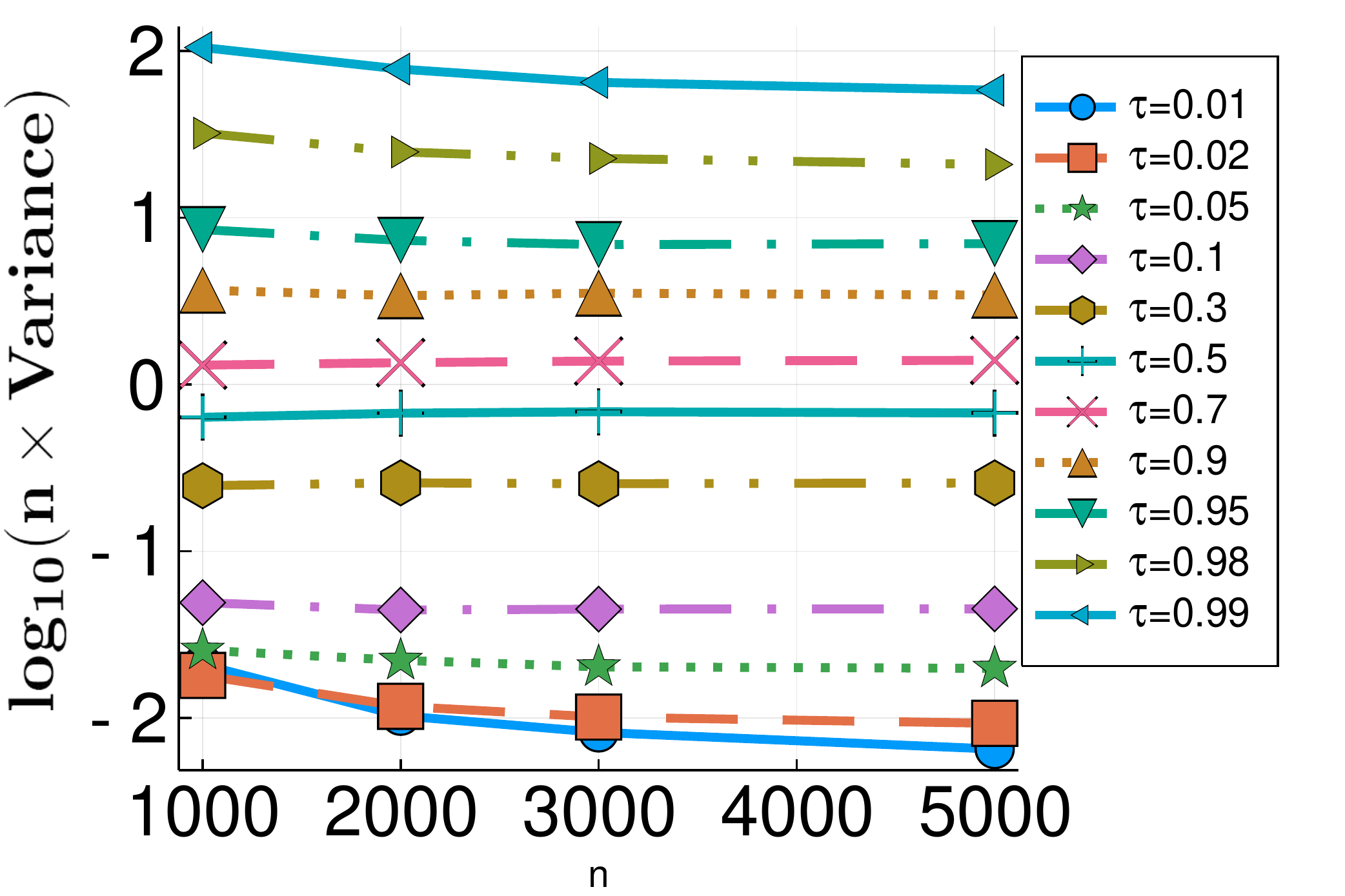}
\caption{Absolute bias and $n\times$variance against relatively large
  values of $n$ with $B=10$ and different quantile levels. Logarithm
  is taken for better presentation. Here, $\X\sim t_3$ and the
  conditional distributions of $Y$ is exponential.}  
  \label{fig:s2}
\end{figure}

\begin{figure}[H]
  \centering
\includegraphics[width=0.485\textwidth]{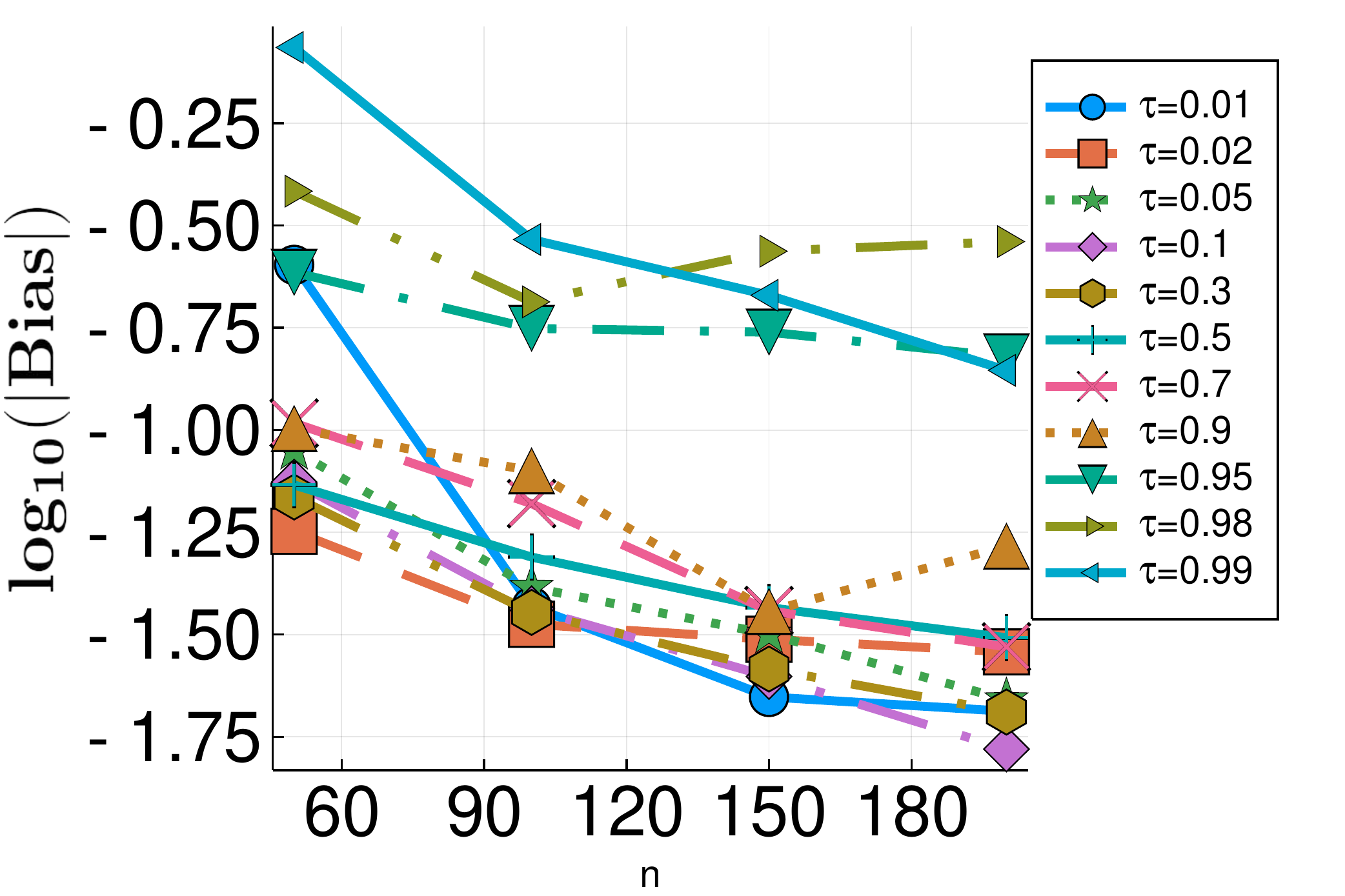}
\includegraphics[width=0.485\textwidth]{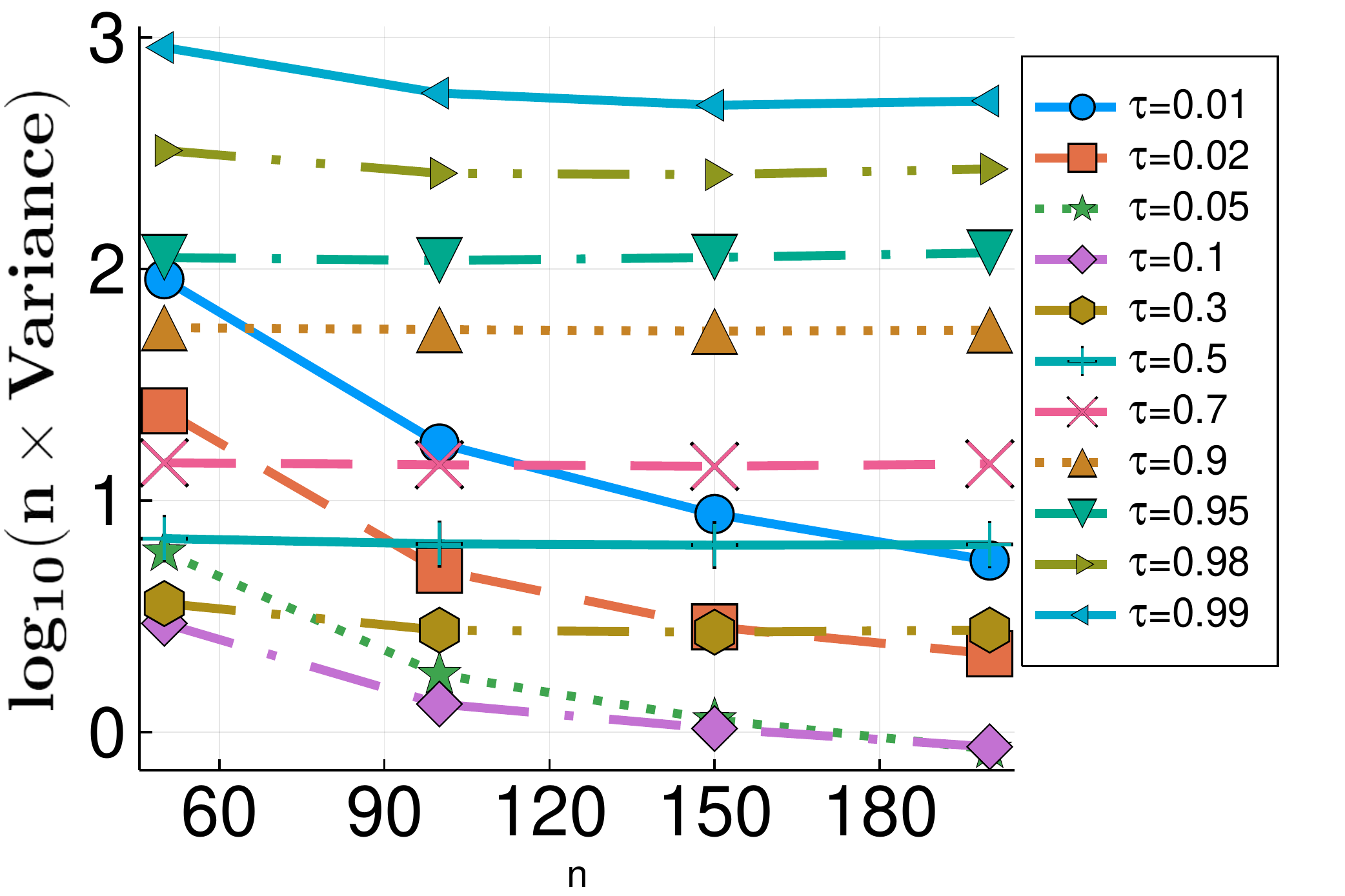}
\caption{Absolute bias and $n\times$variance against small values of
  $n$ with $B=1$ and different quantile levels. Logarithm is taken for
  better presentation. Here, $\X\sim t_3$ and the conditional
  distributions of $Y$ is exponential.}  
  \label{fig:s3}
\end{figure}

Figure \ref{fig:s2} provides the sensitivity of bias and variance to
the subsample size $n$ when $n$ is relatively large. From
Theorem~\ref{thm:4}, we know that when the subsample size is large,
the variance decreases at the $n^{-1}$ rate, so we plot
$n\times$variance against $n$, where the variance is the sum of
variances for all regression coefficients. The exact convergence of
the bias is unknown so we plot the sum of the absolute biases for all
regression coefficients. From Figure \ref{fig:s2}, we see that the
bias has a clear decreasing trend 
when sample size increases, and the variance is clearly decreasing at the
$n^{-1}$ rate for most quantile levels because the curves are
relatively flat.  For extreme quantile levels such as $\tau=0.01,
0.02$, and $0.99$, there is a decreasing pattern for
$n\times$variance, meaning that a larger sample size is required for
the asymptotic distribution to be precise.
Figure \ref{fig:s3} provides similar sensitivity analysis results when $n$ is small. The general trend is
the same, in that both the bias and the variance decrease when $n$
increases. Interestingly, even though the sample sizes are
small, at most quantile levels, we can still see the decreasing of the
variance at the $n^{-1}$ rate. 

\subsection{Universal subsampling probabilities for multiple quantile levels.}\label{sec:manytaus}

\begin{figure}[H]
  \centering
\includegraphics[width=0.45\textwidth,page= 1]{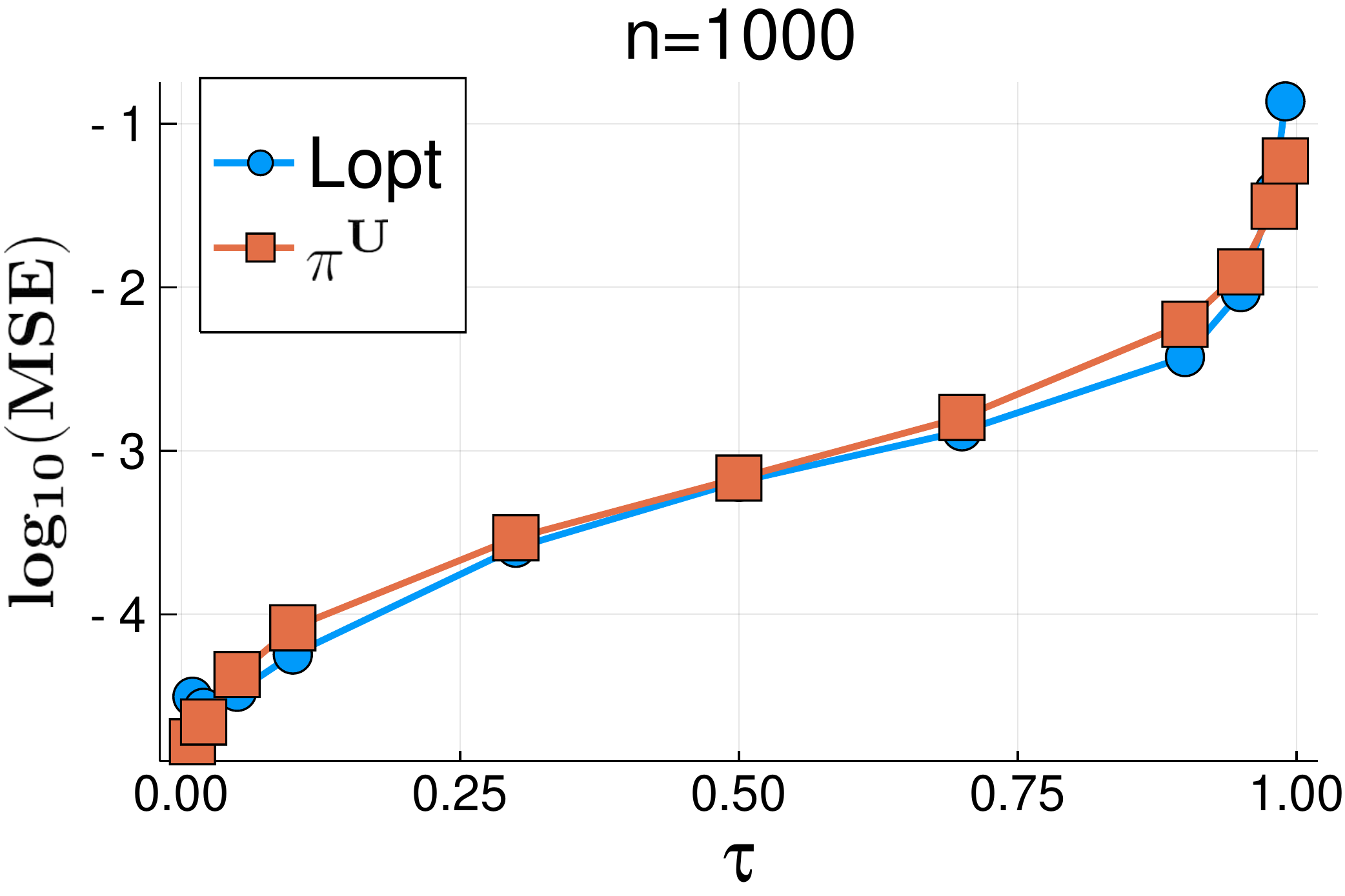}
\includegraphics[width=0.45\textwidth,page= 2]{00msetau.pdf}
\\[3mm]
\includegraphics[width=0.45\textwidth,page= 3]{00msetau.pdf}
\includegraphics[width=0.45\textwidth,page= 4]{00msetau.pdf}
\caption{log$_{10}$(MSE) against against
  quantile level $\tau$ for different subsample size $n$ and a fixed
  $B=10$. $\X\sim t_3$ and the conditional distributions of $Y$ is exponential.}
  \label{fig:s4}
\end{figure}

In this section, we provide additional numerical results to
evaluate the performance of the sub-optimal universal sampling probabilities
$\pi_i^{U}$'s derived at the end of 
Section   \ref{sec:iterative-sampling} in the main paper. We use the
same model setup and sample size configurations as presented in
Section~\ref{sec:mult-quant-levels}. Figure~\ref{fig:s4} presents the
MSE for subsampling estimator based on both  $\pi_i^{\mvc}$ and
$\pi_i^{U}$. We see that although $\pi_i^{U}$ may not be as efficient
as $\pi_i^{\mvc}$ for most quantile levels, the efficiency loss is not
severe.

\newpage
\subsection{Relation between confidence interval and choice of $B$}\label{sec:nB}

In this section, we create confidence intervals and evaluate the
proposed method in terms of the empirical coverage probability. We
first use the same model set up and sample size configurations as in
Section~\ref{sec:numer-exper} of the main paper. Table~\ref{tab:s1}
provides the corresponding results. We see that most of the empirical
coverage probabilities are close to the nominal level of 0.95. Only
when $B=10$, some empirical coverage probabilities may be lower than
0.95. 

To further investigate the scenario that $B$ is relatively large
compared with $n$, we set $n=100$ and set $B=10, 20, 50, 100$, and
$500$. Results are presented in Table~\ref{tab:s2}. It is seen that
the empirical coverage probabilities for the case of $\tau=0.75$ drop
significantly. This indicates that $B$ should be much smaller compared
with $n$ in order to obtain valid inference, which agree with our
theoretical requirement in Section~\ref{sec:iterative-sampling} of the
main paper. This also echos the results in the divide and conquer
literature that the number of partitions should be much smaller than
the sample size in each data partition
\citep[e.g.,][]{schifano2016online, cheng2015computational,
  battey2018distributed,volgushev2017distributed}. Note that the proposed method produces good
results with $B=100$ and $500$ when $\tau=0.5$ in
Table~\ref{tab:s2}. However, this should not be interpreted as that
the proposed method is valid with $n\ge B$.  In fact, we do not know the
asymptotic distribution for this scenario, and the results here may
happen by chance.

\begin{table}[H]
  \caption{Coverage probabilities of 95\% confidence intervals for
    regression coefficients with different values of $B$ and $\tau$
    when $N=10^6$ 
    and $n_0=n=1000$. $\X\sim t_3$ and the conditional distributions of $Y$ is exponential.}
  \label{tab:s1}
 \centering
\begin{tabular}{lcccclcccc}\hline
  & \multicolumn{4}{c}{$\tau=0.5$} &  & \multicolumn{4}{c}{$\tau=0.75$} \\
  \cline{2-5}\cline{7-10}
 & $B=10$ & $B=20$ & $B=50$ & $B=100$ &  
 & $B=10$ & $B=20$ & $B=50$ & $B=100$ \\ \cline{2-10}
$\beta_1$ & 0.931 & 0.927 & 0.941 & 0.950 &  & 0.923 & 0.943 & 0.943 & 0.948 \\ 
$\beta_2$ & 0.940 & 0.938 & 0.940 & 0.940 &  & 0.931 & 0.940 & 0.944 & 0.937 \\ 
$\beta_3$ & 0.936 & 0.957 & 0.939 & 0.949 &  & 0.946 & 0.932 & 0.934 & 0.936 \\ 
$\beta_4$ & 0.924 & 0.941 & 0.947 & 0.951 &  & 0.944 & 0.940 & 0.945 & 0.944 \\ 
$\beta_5$ & 0.914 & 0.938 & 0.935 & 0.952 &  & 0.929 & 0.936 & 0.930 & 0.941 \\ 
$\beta_6$ & 0.949 & 0.937 & 0.931 & 0.938 &  & 0.939 & 0.939 & 0.937 & 0.933 \\ 
  \hline  
\end{tabular}
\end{table}

\begin{table}[H]
  \caption{Coverage probabilities of 95\% confidence intervals for
    regression coefficients with different values of $B$ and $\tau$
    when $N=10^6$ 
    and $n_0=n=100$. $\X\sim t_3$ and the conditional distributions of $Y$ is exponential.} 
  \label{tab:s2}
 \centering
\begin{tabular}{lccccclccccc}\hline
  & \multicolumn{5}{c}{$\tau=0.5$} &  & \multicolumn{5}{c}{$\tau=0.75$} \\
  \cline{2-6}\cline{8-12}
$B=$ & $10$ & $20$ & $50$ & $100$ & $500$ &  
& $10$ & $20$ & $50$ & $100$ & $500$ \\ \cline{2-12}
$\beta_1$ & 0.916 & 0.950 & 0.949 & 0.927 & 0.954 &  & 0.935 & 0.927 & 0.950 & 0.918 & 0.830 \\ 
$\beta_2$ & 0.942 & 0.933 & 0.951 & 0.936 & 0.944 &  & 0.932 & 0.948 & 0.930 & 0.919 & 0.844 \\ 
$\beta_3$ & 0.932 & 0.938 & 0.954 & 0.945 & 0.952 &  & 0.928 & 0.941 & 0.917 & 0.938 & 0.832 \\ 
$\beta_4$ & 0.936 & 0.941 & 0.938 & 0.945 & 0.945 &  & 0.919 & 0.938 & 0.934 & 0.923 & 0.835 \\ 
$\beta_5$ & 0.937 & 0.934 & 0.954 & 0.955 & 0.954 &  & 0.933 & 0.946 & 0.947 & 0.924 & 0.838 \\ 
$\beta_6$ & 0.926 & 0.945 & 0.949 & 0.950 & 0.945 &  & 0.923 & 0.949 & 0.942 & 0.925 & 0.826 \\ 
  \hline  
\end{tabular}
\end{table}

\newpage

\subsection{Computational time}

We provide additional results in terms of computational time and
compare the performance of the proposed method with that of the divide
and conquer method. 

\begin{figure}[H]
  \centering
\includegraphics[width=0.49\textwidth,page= 1]{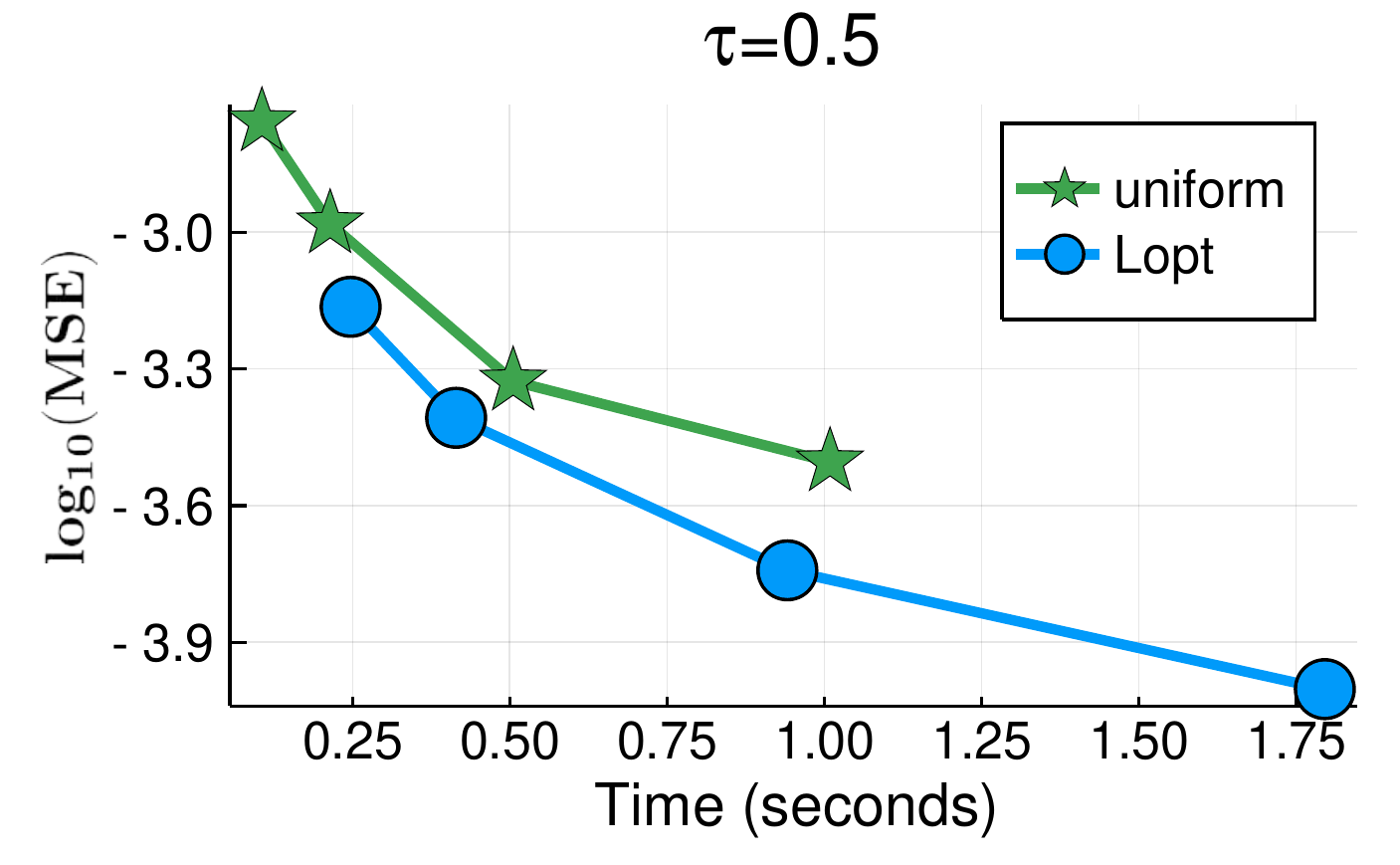}
\includegraphics[width=0.49\textwidth,page= 4]{00time.pdf}
\caption{Empirical MSE vs CPU time (seconds) with $n=1000$ and
  different values of $B$. Here $\X\sim t_2$ and the distribution of
  $Y$ is exponential.
} 
  \label{fig:time1}
\end{figure}

We first plot the MSE against the CPU time (in seconds) for the
proposed method based on both Lopt subsampling and uniform
subsampling. The CPU time is recorded as the average time of ten
repetitions of different methods. In each repetition, we recalculate
the optimal subsampling probabilities so that this overhead time is
taken into account. 
The model set up and sample size configurations are the same as
Section~\ref{sec:numer-exper} of the main paper. It is seen from
Figure~\ref{fig:time1} that the MSE decreases as the CPU time
increases.

\begin{figure}[H]
  \centering
\includegraphics[width=0.49\textwidth,page= 3]{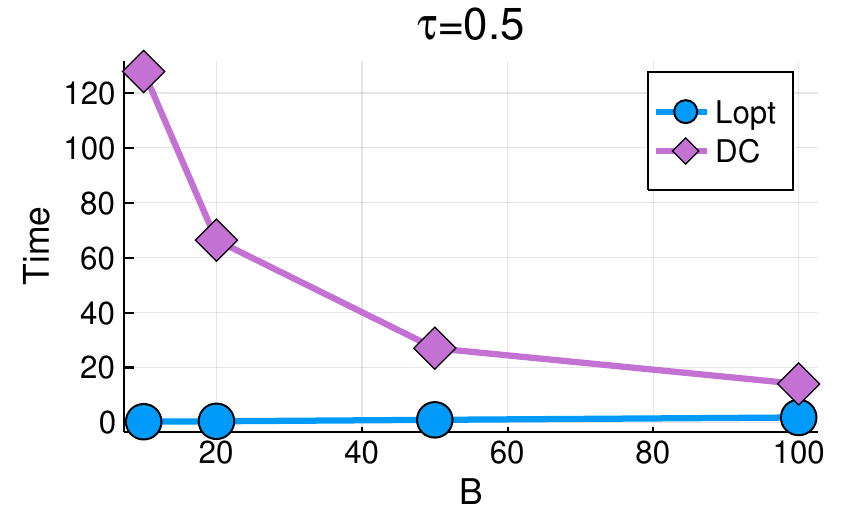}
\includegraphics[width=0.49\textwidth,page= 6]{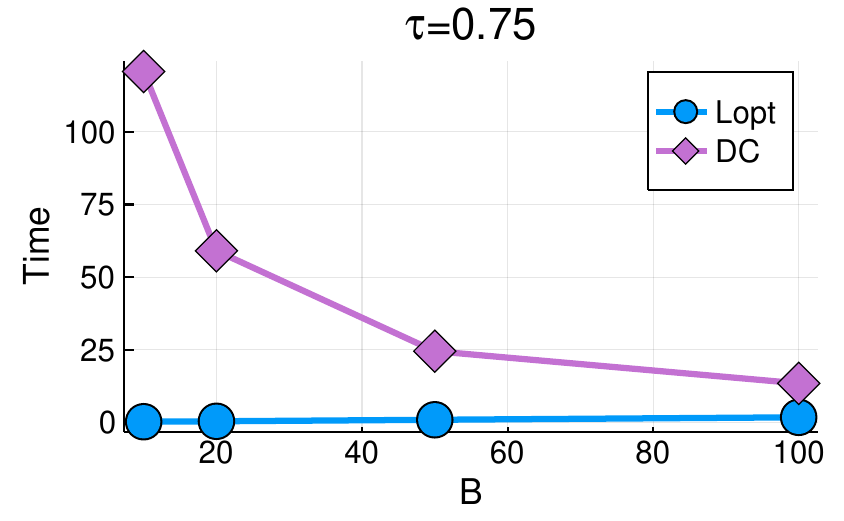}
\caption{CPU time (seconds) vs $B$ with $n=1000$. Here $\X\sim t_2$
  and the distribution of $Y$ is exponential. 
} 
  \label{fig:time2}
\end{figure}

Now we carry out additional numerical experiment to further compare
the computation time of our proposal with that of the divide and
conquer method. For the divide and conquer method,
we divide the full data into $B$ blocks with equal
  number of observations and obtain the estimate from each block of
  data. Let these estimates be $\wh\bbeta_b$ for $b=1, ..., B$.
 We then form the divide and conquer estimator via
  \begin{equation*}
    \wh\bbeta_{DC}=\frac{1}{B}\sum_{b=1}^B\wh\bbeta_b.
  \end{equation*}
Figure \ref{fig:time2} plots CPU times against $B$. Interestingly, we
find that our proposal is much faster than divide 
and conquer method. This shows that even though there is overhead
involved in our method, it is still computationally much less
demanding than the divide and conquer method. 

\begin{figure}[H]
  \centering
\includegraphics[width=0.49\textwidth,page= 2]{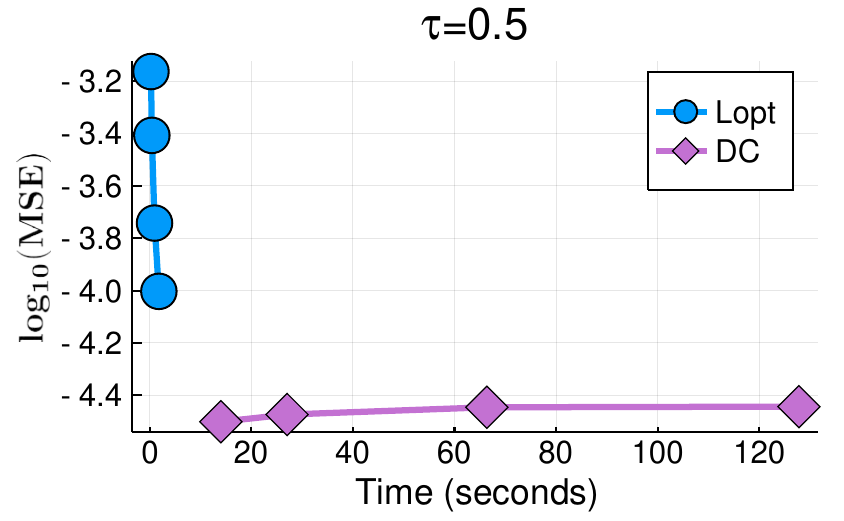}
\includegraphics[width=0.49\textwidth,page= 5]{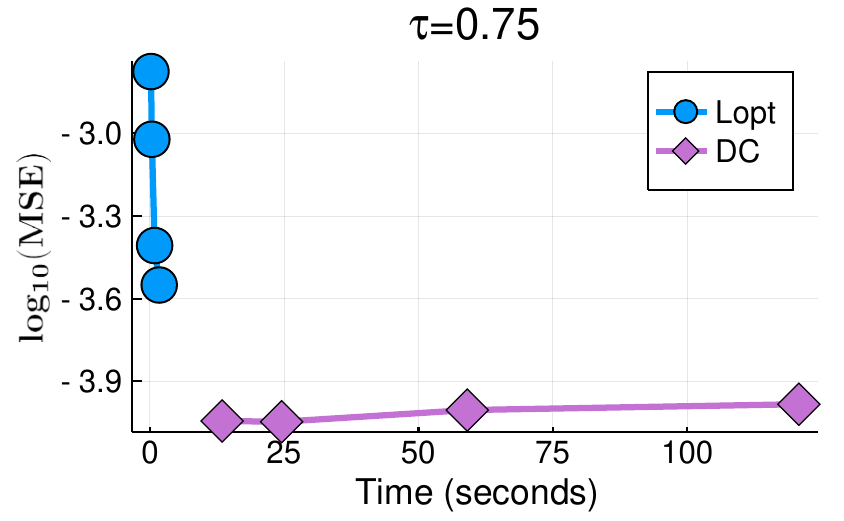}
\caption{Empirical MSE vs CPU time (seconds) with $n=1000$ and
  different values of $B$. Here $\X\sim t_2$ and the distribution of
  $Y$ is exponential. 
} 
  \label{fig:time3}
\end{figure}

Note that the divide and conquer method uses the full data, while our
method is based on a subsample, hence the additional
computational time of the divide and conquer method also brings gain
in terms of MSE. This is similar to the fact that MSE based on the
full data is much smaller than that based on a subsample. To further
illustrate this fact, we plotted the MSE as a function of computation
time in Figure \ref{fig:time3}. We see that the two
methods occupy different regions in the plots, indicating that the
computation times of the two methods are very different and their
estimation precisions are also very different. When both computation
and precision are taken into account, there is no clear winner. Hence
which method is more applicable depends on the practical needs.

\end{document}